\documentclass[twocolumn,secnumarabic,amssymb, nobibnotes, prb, supebrscriptaddress]{revtex4-2}

\usepackage{xcolor}
\usepackage{amsmath}
\usepackage{physics}
\usepackage{graphicx}
\usepackage{wrapfig}
\usepackage{float}
\usepackage{booktabs}
\usepackage{multirow}
\usepackage{array}
\usepackage{setspace}
\graphicspath{{Figs/}}

\begin{document}

\title{Role of chemical disorder in tuning the Weyl points in vanadium doped  Co$_2$TiSn}

\author{Payal Chaudhary}
\affiliation{School of Materials Science and Technology, Indian Institute of Technology (Banaras Hindu University), Varanasi 
221005, India}
\author{Krishna Kant Dubey}
\affiliation{School of Materials Science and Technology, Indian Institute of Technology (Banaras Hindu University), Varanasi 
221005, India}
\author{Gaurav K. Shukla}
\affiliation{School of Materials Science and Technology, Indian Institute of Technology (Banaras Hindu University), Varanasi 
221005, India}
\author{Surasree Sadhukhan}
\affiliation{School of Physical Sciences, Indian Institute of Technology Goa, Goa 403401, India}
\author{Sudipta Kanungo}
\affiliation{School of Physical Sciences, Indian Institute of Technology Goa, Goa 403401, India}
\author{Ajit K. Jena}
\affiliation{Indo-Korea Science and Technology Center (IKST), Bangalore 560065, India}
\author{S.-C Lee}
\affiliation{Indo-Korea Science and Technology Center (IKST), Bangalore 560065, India}
\author{S. Bhattacharjee}
\affiliation{Indo-Korea Science and Technology Center (IKST), Bangalore 560065, India}
\author{Jan Min\'ar}
\affiliation{New Technologies Research Centre, University of West Bohemia, Univerzitní 8, CZ-306 14 Pilsen, Czech Republic}
\author{Sanjay Singh*}
\affiliation{School of Materials Science and Technology, Indian Institute of Technology (Banaras Hindu University), Varanasi 
221005, India}
\author{Sunil Wilfred D'Souza,}
\affiliation{New Technologies Research Centre, University of West Bohemia, Univerzitní 8, CZ-306 14 Pilsen, Czech Republic}

\begin{abstract}
The  lack of time-reversal symmetry and Weyl fermions give exotic transport properties to Co-based Heusler alloys. In the present 
study, we have investigated the role of chemical disorder on the variation of Weyl points in  Co\textsubscript{2}Ti
\textsubscript{1-x}V\textsubscript{x}Sn magnetic Weyl semimetal candidate. We employ the first principle approach to track the 
evolution of the nodal lines responsible for the appearance of Weyl node in Co$_2$TiSn as a function of V substitution in place of 
Ti. By increasing the V concentration in place of Ti, the nodal line moves toward Fermi level and remains at Fermi level around 
the middle composition. Further increase of the V content, leads shifting of nodal line away from Fermi level. Density of state 
calculation shows half-metallic behavior for the entire range of composition. The magnetic moment on each Co atom as a function of 
V concentration increases linearly up to x=0.4, and after that, it starts decreasing. We also investigated the evolution of the Weyl nodes and Fermi arcs with chemical doping. The first-principles calculations reveal that via replacing almost half of the Ti with V, the intrinsic anomalous Hall conductivity increased twice as compared to the undoped composition. Our results indicate that the composition close to the 50\% V doped Co$_2$TiSn, will be an ideal composition for the experimental investigation of Weyl physics.  
\end{abstract}

\maketitle

\section{Introduction}
Weyl semimetals (WSMs) have created vast interest in recent years due to their novel electronic and transport properties \cite{hu2019transport,RevModPhys.90.015001}, such as very high electron mobilities \cite{shekhar2015extremely}, Fermi arcs on the surface \cite{wan2011topological,jia2016weyl}, extremely large magnetoresistance \cite{son2013chiral,shekhar2015extremely}, anomalous Hall effect \cite{burkov2014anomalous,Shekhar9140}, and the anomalous Nernst effect \cite{ikhlas2017large,sakai2018giant}. WSMs also exhibit unconventional optical properties, such as large and quantized photo-currents \cite{taguchi2016photovoltaic,chan2017photocurrents,de2017quantized,osterhoudt2017colossal}, second-harmonic generation \cite{morimoto2016topological,wu2017giant}, and Kerr rotation \cite{feng2015large,higo2018large}. These properties can lead to more efficient electronic and photonic applications \cite{hu2019transport}. WSMs are an especial class of the topological materials, characterized by the crossings of singly degenerate energy bands near the Fermi energy, leading to the formation of pairs of Weyl nodes \cite{Xu15,Xu2015,hasan17}. WSMs provide a platform for manipulating and understanding the physics of the chiral Weyl fermions \cite{huang2015observation,yan2017topological}.

Inversion symmetry (IS) or time-reversal symmetry (TRS) must be broken to obtain Weyl nodes/WSMs \cite{RevModPhys.90.015001,yan2017topological,AJR2P}. WSMs with broken IS have been investigated extensively \cite{Xu15,Xu2015,shekhar2015extremely}, while WSMs with broken TRS, known as magnetic WSMs, were recently discovered in experiments\cite{ikhlas2017large,belopolski2019discovery}. Magnetic WSMs created much interest because, in this class of WSMs, the properties can be manipulated using a magnetic field as an external degree of freedom. Heusler alloys have emerged as an important class of materials to investigate the Weyl physics and its consequences \cite{liu2017nonmagnetic,belopolski2019discovery,manna2018colossal,ernst2019anomalous,dulal2019weak,nakajima2015topological,kim2018beyond,Shekhar9140}. In Heusler compounds, we either look at half-Heusler or inverse Heusler compounds (IS breaking) \cite{liu2017nonmagnetic}, at magnetic compounds (TRS breaking) \cite{belopolski2019discovery,manna2018colossal,ernst2019anomalous,dulal2019weak}, or compounds with both IS and TRS breaking \cite{nakajima2015topological,kim2018beyond,Shekhar9140}. In most magnetic Heuslers, the magnetization direction can be changed quite easily. Since the location of Weyl nodes in the momentum space depends on the direction of magnetization, Heusler compounds can prove to be useful to understand the physics of Weyl fermions. Combined with their extensive tunability, Heusler WSMs are a promising platform for practical topological applications \cite{hu2019transport,RevModPhys.90.015001}.

Co\textsubscript{2}MnGa has been theoretically predicted and experimentally proven to be a WSM. The Weyl nodes lie close to the Fermi energy \cite{belopolski2019discovery,chang2017topological,belopolski2019discovery}, and transport measurements have shown large anomalous Hall and Nernst effects \cite{sakai2018giant,reichlova2018large,guin2019anomalous,takashi2019signs,markou2019thickness,park2020thickness}.
Besides Co$_2$MnGa, other Heusler compounds have also been predicted to be WSM \cite{kubler2016weyl,chang2016room,wang2016time,chadov2017stability}. Although, for most of the proposed Heusler WSMs, the Weyl nodes lie away from the Fermi energy, which reduces the topological properties of these materials \cite{kubler2016weyl,chang2016room,wang2016time,ernst2019anomalous,chadov2017stability,manna2018colossal}. By tuning the Fermi energy, it can coincide with the energy of the nodes, which can significantly improve the properties \cite{wang2016time,kushwaha2018magnetic,yang2019magnetic}.

Co\textsubscript{2}TiSn, a Heusler compound that has a high Curie temperature and shows half-metallic behavior \cite{barth2010itinerant,barth2011anomalous,ooka2016magnetization,bainsla2016spin,shigeta2018pressure}, has been proposed as a WSM candidate \cite{chang2016room,wang2016time,ernst2019anomalous}. Co\textsubscript{2}TiSn has 26 valence electrons and has Weyl nodes with chemical potential a few hundred meV above the Fermi energy \cite{chang2016room,wang2016time,ernst2019anomalous}. 
~The number of valence electrons must be increased to make the nodes chemical potential coincide with the Fermi energy \cite{ernst2019anomalous}. To achieve this, Ti, which has 2 electrons in its valence 3\textit{d} orbital, can be substituted with V, which has 3 electrons in the 3\textit{d} orbital. Wang et al. suggested that the doping of 10\% V in place of Ti (i.e., Co$_2$Ti$_{0.9}$V$_{0.1}$Ga)  to obtain the WSM phase \cite{wang2016time}. However, efforts to synthesize thin films Co$_2$Ti$_{0.9}$V$_{0.1}$Sn have not been successful, and the most stable composition was found to be Co$_2$Ti$_{0.6}$V$_{0.4}$Sn (i.e., 40\% V doped) \cite{hu2019unconventional}. Transport measurements for Co\textsubscript{2}TiSn and Co\textsubscript{2}Ti\textsubscript{0.6}V\textsubscript{0.4}Sn thin films show anomalous Hall and Nernst effects \cite{hu2018anomalous,hu2019unconventional}. Interestingly, the anomalous Nernst angle and coefficient for the doped compound (i.e., Co$_2$Ti$_{0.6}$V$_{0.4}$Sn) were significantly higher than the undoped one \cite{hu2018anomalous}. 

In this manuscript, we investigated the effect of V doping on the electronic structure of Co\textsubscript{2}Ti\textsubscript{1-x}V\textsubscript{x}Sn (x = 0.0, 0.2, 0.4, 0.6, 0.8 and 1.0). The position of nodal lines, responsible for Weyl node formation in Co$_2$TiSn, changes as a function of V substitution in place of Ti. With increasing V in place of Ti, nodal lines and Fermi arcs get shifted towards the Fermi level close to the mid composition (i.e Co$_2$Ti$_{0.6}$V$_{0.4}$Sn). The intrinsic anomalous Hall conductivity obtained from the theory for the 50\% V doped composition (Co$_2$Ti$_{0.5}$V$_{0.5}$Sn) is nearly twice as compared to the undoped composition, due to the existence of the Fermi arcs very close ($\approx$ 7-23 meV) to the Fermi energy. 
\par

\section{Methods}
The calculations were performed using the full-potential Korringa$–$Kohn$–$Rostoker (KKR) Green's function method, as implemented in the SPRKKR package \cite{ebert2005munich} and pseudo-potential based density functional theory (DFT) as implemented in Quantum ESPRESSO (QE)\cite{giannozzi2009quantum} and Vienna ab-initio simulation package (VASP)\cite{kresse1993ab, kresse1996efficient}. The exchange-correlation potential is approximated through PBE-GGA functional \cite{perdew1996generalized}. A k-mesh consisting of $(22 \times 22 \times 22)$ k-points was used, and the angular momentum cut-off number was chosen to be $l_{max} = 3$. The Fermi energies were determined using the Lloyd formula \cite{ebert2011calculating}. For the calculations of the Bloch spectral functions  spin-orbit coupling (SOC) has been taken into account. The band structures were calculated for both the spin-polarized, non spin-orbit coupling case as well as the case with spin-orbit coupling, taken as the a scalar relativistic correction to the original Hamiltonian. The ground state energies obtained when the magnetization quantization directions are kept in the [001], [110] and [111] directions are found to be the same within the limits of the methodology, indicating that the  magnetization directions, and therefore the positions of the Weyl nodes, can be changed easily. This matches with the literature on Co$_2$TiSn, Co$_2$VSn and similar Heusler compounds \cite{chang2016room,wang2016time}. The disorder was taken into account through the coherent potential approximation (CPA) in the SPR-KKR calculations \cite{soven1967coherent,soven1969contribution}. Optimized pseudopotentials \cite{hamann2013optimized} are used in the calculations and the kinetic energy cutoff for the planewave is taken as 80 $Ry$. The electronic integration over the Brillouin zone (BZ) is approximated by the Gaussian smearing of 0.005 $Ry$ both for the self-consistent (sc) and non-self-consistent (nsc) calculations. The Monkhorst-Pack \textbf{k}-grid of $8\times8\times8$ are considered for the Brillouin zone integration for the DFT band structure calculations. The Wannier interpolated bands, the anomalous Hall conductivity (AHC), normalized Berry curvature and the Fermi arcs were calculated using Wannier90 \cite{marzari1997maximally, souza2001maximally, pizzi2020wannier90} and WannierTool \cite{wu2018wanniertools} starting from the the plane wave based pseudo potential DFT band structures. The transition metal $d$-orbitals are used in the energy selective downfolding projections for the wannier90 calculations. The AHC calculation is carried out with a dense \textbf{k}-grid of $75\times75\times75$. Further, through the  adaptive refinement technique a fine mesh of $5\times5\times5$ is added around the points wherever the mod of the Berry curvature ($\abs{\Omega(\textbf{k})}$) exceeds 100 Bohr$^2$. All the calculated structures are optimized with tight convergence threshold both for the  energy ($10^{-10} Ry$) and Feynman Hellman force ($10^{-10} Ry/Bohr$). The self consistent calculations are converged with the energy cut-off of $10^{-8} Ry$.

\par

\section{Results and Discussion}
\subsection{Optimisation of lattice parameters}

\begin{figure}[t]
    \includegraphics[width=\linewidth]{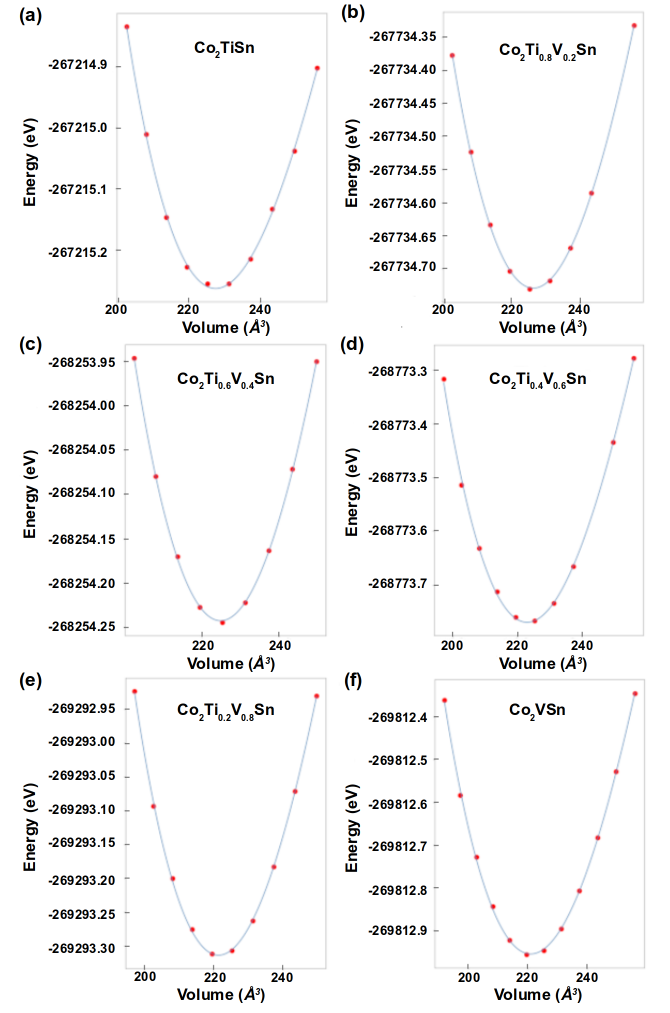}
    \caption{Total energy vs. volume of the unit cell for Co$_2$Ti$_{1-x}$V$_{x}$Sn (x = 0.0, 0.2, 0.4, 0.6, 0.8, 1.0). The Birch-Murnaghan equation of state is used to determine the equilibrium values \cite{birch1947finite}.}
    \label{fig:figure-1}
\end{figure}

The lattice parameters for all the compositions were obtained by varying the parameters and calculating the respective ground state energies. The equilibrium  values were found using the Birch-Murnaghan equation of state \cite{birch1947finite} fit for the total energy as a function of the unit cell volume. The plots of total energy vs. the unit cell volume are shown in Fig. \ref{fig:figure-1}. The calculated lattice parameters values, 6.104 \AA (for x=0.0), 6.087 \AA (for x=0.2), 6.083 \AA (for x=0.4), 6.065 \AA (for x=0.6), 6.050 \AA (for x=0.8)~and 6.04 \AA (for x=1.0) are in well agreement with the experimentally reported lattice parameters 6.076\AA, 6.051 \AA,  6.040 \AA,   6.034\AA,  6.014\AA,  5.98 \AA~for x = 0.0, 0.2, 0.4, 0.6, 0.8, and 1.0 respectively, which follow similar trend with change in the composition\cite{dunlap1982conduction,pendl1996investigation}.
\par

\begin{figure}[htbp]
    \includegraphics[width=1\linewidth]{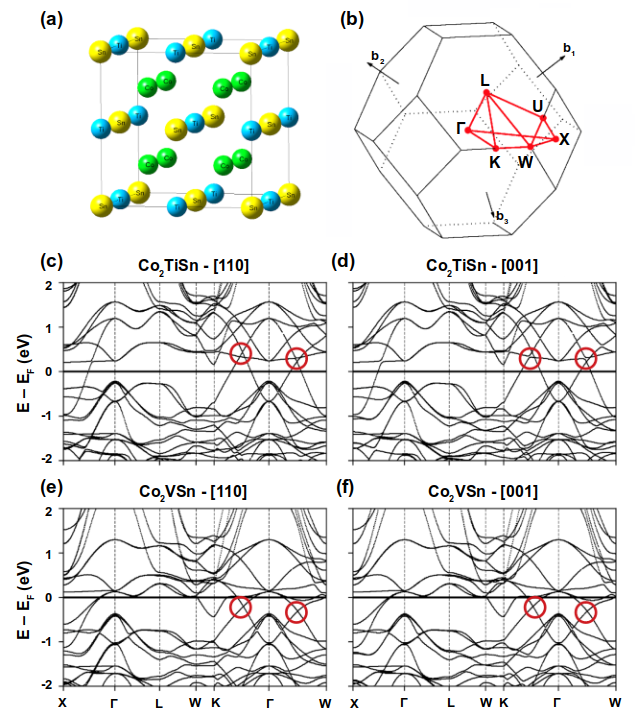}
\caption{(a)Crystal structure of Co\textsubscript{2}TiSn. Co atoms are represented by green spheres, Ti atoms by blue spheres, and Sn atoms by yellow spheres. (b) Brillouin zone showing the high-symmetry points and the k-path followed in the band structures for the primitive unit cell. (c),(e) Calculated band structures of Co\textsubscript{2}TiSn and Co\textsubscript{2}VSn with spin-orbit coupling in the [110] quantization direction. (d),(f)Calculated band structures of Co\textsubscript{2}TiSn and Co\textsubscript{2}VSn with spin-orbit coupling in the [001] quantization direction. The red circles indicate the location of the crossings in the Co-Y (Y = Ti, V) hybridized 3d bands.}
 \label{fig:figure-2}
\end{figure}

\subsection{Band structure calculations for stoichiometric Co\textsubscript{2}TiSn and Co\textsubscript{2}VSn Heusler compounds}

The band structures calculated using plane wave based pseudo-potential for the end (stoichiometric) compositions, with spin-orbit coupling (SOC), are shown in Fig. \ref{fig:figure-2}. The band structures for both the stoichiometric compositions, Co$_2$TiSn and Co$_2$VSn, have been calculated using the  Heusler (L2$_1$) cubic structure with space group $Fm\Bar{3}m$. In the conventional unit cell of this structure, Co occupies the 8c (1/4, 1/4, 1/4) Wyckoff position, Ti  (or V) occupies 4b (1/2, 1/2, 1/2), and Sn is at 4a (0, 0, 0), as shown in Fig. \ref{fig:figure-2}(a) for Co$_2$TiSn as an example.

\begin{figure*}[t]
    \includegraphics[width=\linewidth]{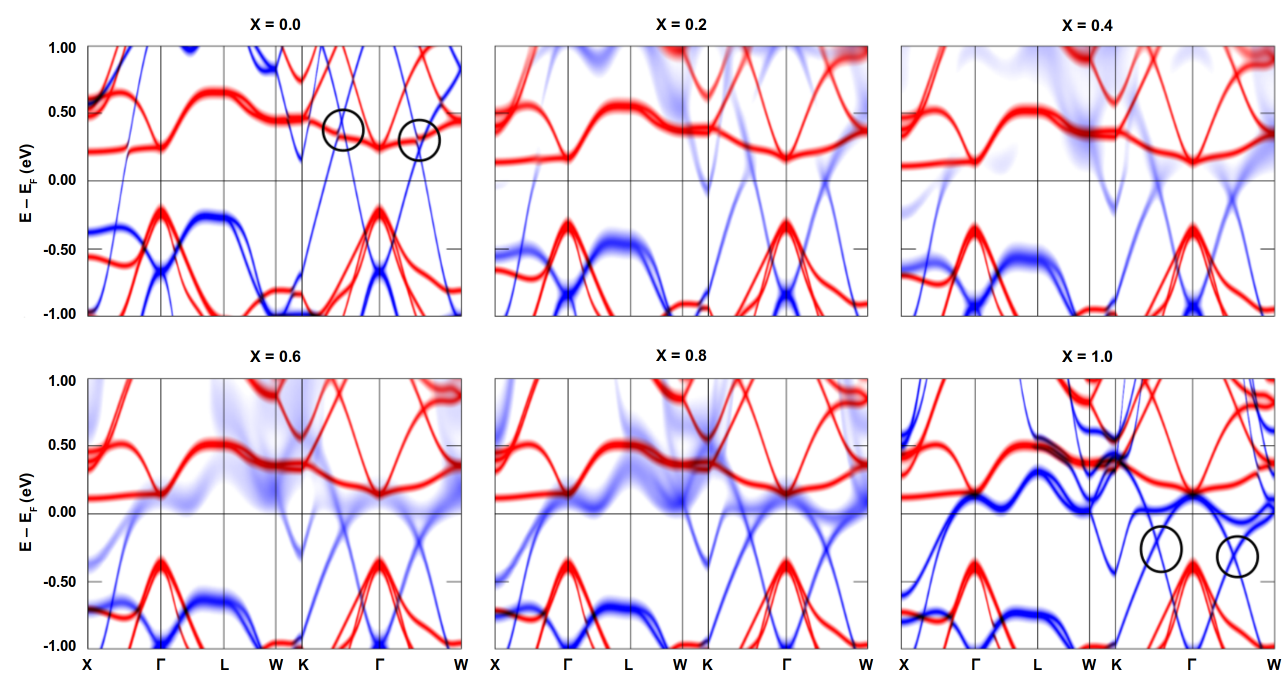}
    \caption{Bloch spectral function plots of Co\textsubscript{2}Ti\textsubscript{1-x}V\textsubscript{x}Sn for the primitive unit cell. The majority and minority spin states are represented by the blue and red lines, respectively. The nodal line can be seen shifting downwards with respect to the Fermi level as the concentration of V increases, tuning with the Fermi level at x = 0.2 and x = 0.4.}
    \label{fig:figure-3}
\end{figure*}

Fig. \ref{fig:figure-2} shows the spin-orbit coupled band structures along the high symmetry lines in the BZ for Co$_2$TiSn and Co$_2$VSn with magnetization oriented along the [110] and [001] directions. Red circles mark the nodal lines of interest which may form the Weyl nodes. An analysis of the hybridisation and symmetry of these nodal lines is detailed in the appendix. In Co$_2$TiSn, the nodal line remains unaffected when the magnetization is oriented in the [001] direction (Fig. \ref{fig:figure-2}(d)), but very slight gaps occur when it is oriented in the [110] direction (Fig. \ref{fig:figure-2}(c)). In the band structure for the [110] magnetization direction (Fig. \ref{fig:figure-2}(c)), the nodal line crossing along $\Gamma-W$ has a very small gap, not visible in the figure, while the crossings along $\Gamma-X$ and $\Gamma-K$ remain unaffected. The crossings of the nodal line bands with the surrounding band along $\Gamma-W$ and $\Gamma-K$ also have gaps, according to the magnetization direction, as can be seen next to the red circles in the figures (Fig. \ref{fig:figure-2}(c) and \ref{fig:figure-2}(d)). In Co$_2$VSn, these nodal lines lie entirely below the Fermi energy, as seen in Fig. \ref{fig:figure-2}(e) and \ref{fig:figure-2}(f).

\subsection{Bloch spectral functions for Co\textsubscript{2}Ti\textsubscript{1-x}V\textsubscript{x}Sn}

For disordered compounds, it is difficult to determine the $E$ versus $\textbf{k}$ dispersion relations using methods for periodic ordered systems, such as the plane wave pseudo potential based DFT calculations. One approach is to construct a supercell to add the substituted element in the required ratio and calculate the dispersion relations. However, this results in complex band structures with additional bands due to symmetry. These supercell bands can then be unfolded to get an effective band structure\cite{popescu2010effective,popescu2012extracting}. Using this approach to calculate the relations for a range of compositions is cumbersome and is feasible only for specific compositions.

Here, we use Bloch spectral functions to represent the electronic structure. The Bloch spectral function $A_{B}(\mathbf{k},E)$, defined as the Fourier transform of the Green's function $G(\mathbf{r},\mathbf{r'},E)$, can be written as

\begin{multline}
    A_{B}(\mathbf{k},E) = - \frac{1}{\pi N} \Im Tr \lbrace  \sum_{n,n'}^{N} e^{i\mathbf{k}(\mathbf{R_{n}}-\mathbf{R_{n'}})} \\ \times \int_{\Omega} d^{3}r G(\mathbf{r}+\mathbf{R_{n}},\mathbf{r}+\mathbf{R_{n'}},E) \rbrace 
\end{multline}

This function can be interpreted as the $\textbf{k}$-resolved density of states \cite{ebert2011calculating}.

The Bloch spectral function (BSF) plots are shown in Fig. \ref{fig:figure-3}. The nodal line can be seen shifting downward with respect to the Fermi energy as the concentration of V increases. At x = 0.4, the point of highest energy of the nodal line tunes with the Fermi energy, along $\Gamma-K$. This has been also seen in the Fermi surface plots (Fig. \ref{fig:figure-4}) where for  $x = 0.4$, the two bands touch at the extremities, along $\Gamma-K$. For higher V doped composition e.g. x= 0.6, the nodal lines lie entirely below the Fermi energy.This means that for all compositions having V concentration in the range of $0.6 \leq$x$ \leq1$, the nodal lines lie entirely below the Fermi level. 

A distinct feature of the BSF plots of the substituted compounds are the broadening of the majority bands. This broadening occurs in the energy range -0.5 eV to 2.0 eV. These are the same Co-X (X = Ti, V) hybridized 3\textit{d} bands which form the nodal lines of interest. In the substituted compounds, the bands in this energy range are primarily formed by the 3\textit{d} states on the 4b Wyckoff position, which contain the Ti and V atoms, and have a negligible contribution from Co atoms. The Co atoms in the substituted compounds have more contribution in the states below the nodal lines. The broadening in the bands occurs due to the randomly substituted additional 3\textit{d} electron of the V atom. Since the band corresponding to the additional electron has a different energy, random fluctuations are induced in the energy window of the 3\textit{d} states. The new electronic state manifests as an intermediate state, which can be seen becoming more well defined as the concentration of V increases. 

\subsection {Fermi surface}

\begin{figure*}[htb]
    \includegraphics[width=1\linewidth]{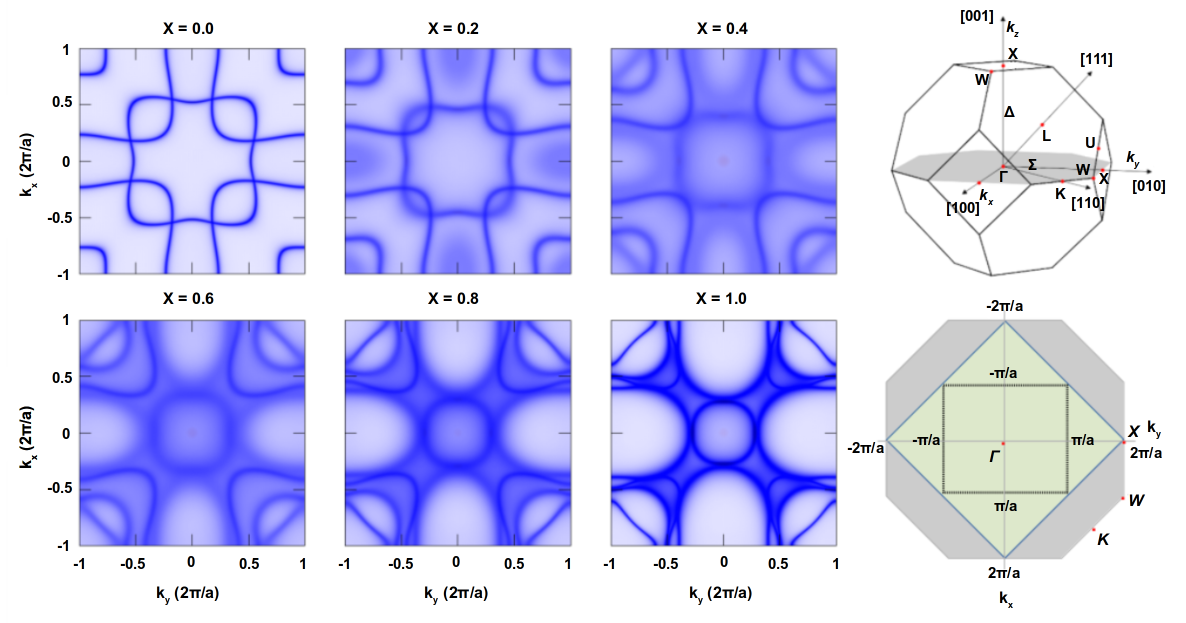}
    \caption{Fermi surface plots of Co\textsubscript{2}Ti\textsubscript{1-x}V\textsubscript{x}Sn with chemical potential set at the Fermi energy for the spin polarized calculations without SOC}. Majority spin states are presented in blue. Since all the compositions are half-metallic, there are no minority spin states at the Fermi level. On the right is a diagram of the Brillouin zone (BZ) in the conventional unit cell setting showing the high-symmetry lines and points, along with the Fermi surface cross-section (cut through the BZ at $k_z = 0$). The dotted shaded square region shows the first BZ.
    \label{fig:figure-4}
\end{figure*}

The evolution of the Fermi surface with respect to the V concentration can be seen in Fig. \ref{fig:figure-4}. These calculations were done by using the primitive unit cell, however, the data is represented in the conventional settings ($k_x$, $k_y$, $k_z$ axis) of the BZ for better presentation. The Fermi surface plots show the electronic states in the $xy$-plane of the BZ, at $k_z = 0$, lying on the Fermi energy. On the right is a schematic diagram showing the high-symmetry points in the Brillouin zone and a cross-section of the Brillouin zone at $k_z = 0$. The intermediate state formed by the addition of the 3\textit{d} V electron can be seen becoming more well-defined along the $X-\Gamma$ and $W-K$ directions as the concentration of V increases. The bands which form the nodal lines can be seen clearly in the Fermi surface plots; one has a Fermi surface around the $\Gamma$ point and the other around the $K$ point. At $x = 0.2$, the nodal line tunes with the Fermi energy along $\Gamma-W$, as can be seen in both the BSF and Fermi surface plots.At $x = 0.4$, the Fermi surface plot shows the two bands touching at the extremities, along $\Gamma-K$. For $0.6 \leq$x $\leq1$, the nodal lines lie entirely below the Fermi energy.
\subsection{Density of states and magnetization}

\begin{figure*}[t]
    \includegraphics[width=\linewidth]{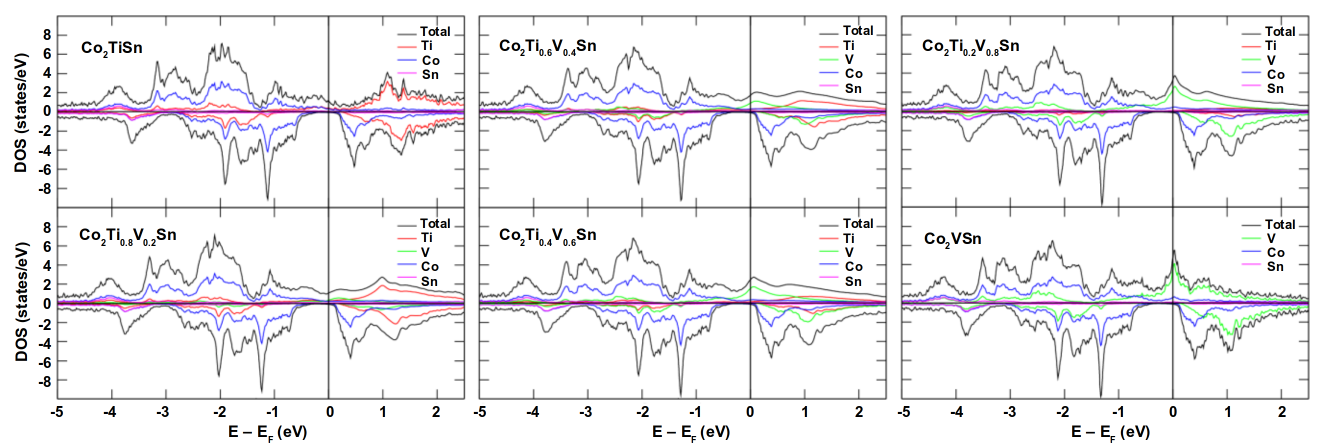}
    \caption{Density of states of Co\textsubscript{2}Ti\textsubscript{1-x}V\textsubscript{x}Sn (x=0.0, 0.2, 0.4, 0.6, 0.8 and 1.0). For each compound, the positive y-axis represents the density of the majority spin states, and the negative y-axis represents the minority spin states.}
    \label{fig:figure-5}
\end{figure*}

The stoichiometric compounds Co$_2$TiSn and Co$_2$VSn have been reported as half-metallic ferromagnets 
\cite{doi:10.1063/1.1853899,hickey2006Fermi,kandpal2007calculated,barth2010itinerant,aguayo2011density}. To investigate the half-
metallic character as a function of chemical disorder, we performed density of states (DOS) calculations as a function of x in Co
\textsubscript{2}Ti\textsubscript{1-x}V\textsubscript{x}Sn. The evolution of the density of states (DOS) is given in Fig. 
\ref{fig:figure-5}. The Fermi energy always lies in the minority spin band gap for all compositions, indicating half-metallic 
behavior throughout. The magnitude of the gap increases slightly with the increase in V concentration, and the gap for  the 
stoichiometric compounds is in good agreement with the literature. The band gap ($\Delta$E) for compositions x=0.0, 0.2, 0.4, 0.6, 
0.8 and 1.0 is found to be 0.491 eV, 0.505 eV, 0.518 eV, 0.518 eV, 0.532 eV and 0.546 eV, respectively. This reveals that the 
$\Delta$E is slightly increasing with increasing V substitution in place of Ti. 

Having obtained the evaluation of nodal lines and half-metallic character, we turn our discussion regarding the magnetic behavior 
of Co\textsubscript{2}Ti\textsubscript{1-x}V\textsubscript{x}Sn alloys. The magnetic moments calculated for all the compounds are 
given in Table \ref{table:table-1}.  Interestingly the magnetic moment of  Co increases initially and after that, it starts 
decreasing, which is in very well agreement with the experimental findings (Table \ref{table:table-1}) \cite{dunlap1982conduction, 
pendl1996investigation}. The magnetic moment of Co is oriented antiparallel with the Ti moment and parallel with the V moment. Ti 
has a low magnetic moment of around 0.1 $\mu_{B}$ per atom, while V has a higher moment ranging from 0.73 to 0.88 $\mu_{B}$ per 
atom. The total magnetic moment per formula unit obtained here shows an increasing trend  with V substitution and found 3$\mu_{B}
$, which is well agreement with previous reported  literature\cite{shukla2020destruction}. Thus our all results are in accordance 
with the observed behavior in the experiments and explain the unusual large anomalous Nernst coefficient for the  Co$_2$Ti$_{0.6}
$V$_{0.4}$Sn as compared to the stoichiometric compound Co$_2$TiSn \cite{hu2018anomalous}. 

\begin{table}[htbp]
    \centering
    \begin{tabular}{ c c c c c c c c c } 
        \toprule
        \multirow{2}{*}{} &
            Experimental ($\mu_{B}$) &
            \multicolumn{7}{c}{Calculated ($\mu_{B}$)} \\
        $x$ & $\mu_{Co}$ \cite{pendl1996investigation} & $\mu_{Co}$ && $\mu_{Ti}$ && $\mu_{V}$ && $\mu_{total}$ \\
        \midrule
        $0.0$ & 0.98 & 1.078 && -0.117 && - && 2.03 \\
        $0.2$ & 0.98 & 1.097 && -0.116 && 0.719 && 2.23 \\
        $0.4$ & 1.07 & 1.097 && -0.126 && 0.830 && 2.43 \\
        $0.6$ & 0.91 & 1.091 && -0.132 && 0.873 && 2.63 \\
        $0.8$ & 0.88 & 1.089 && -0.145 && 0.878 && 2.82 \\
        $1.0$ & 0.60 & 1.094 && - && 0.851 && 3.00 \\
        \bottomrule
    \end{tabular}
    \caption{Magnetic moments for Co\textsubscript{2}Ti\textsubscript{1-x}V\textsubscript{x}Sn (x=0.0, 0.2, 0.4, 0.6, 0.8 and 1.0). The moments for individual atoms are given in units of $\mu_{B}$ per atom, and the total moment is given in $\mu_{B}$ per formula unit.}
    \label{table:table-1}
\end{table}

\subsection{Weyl nodes, Fermi arcs and Berry curvatures}

\begin{figure*}[t]
    \includegraphics[width=0.7\linewidth]{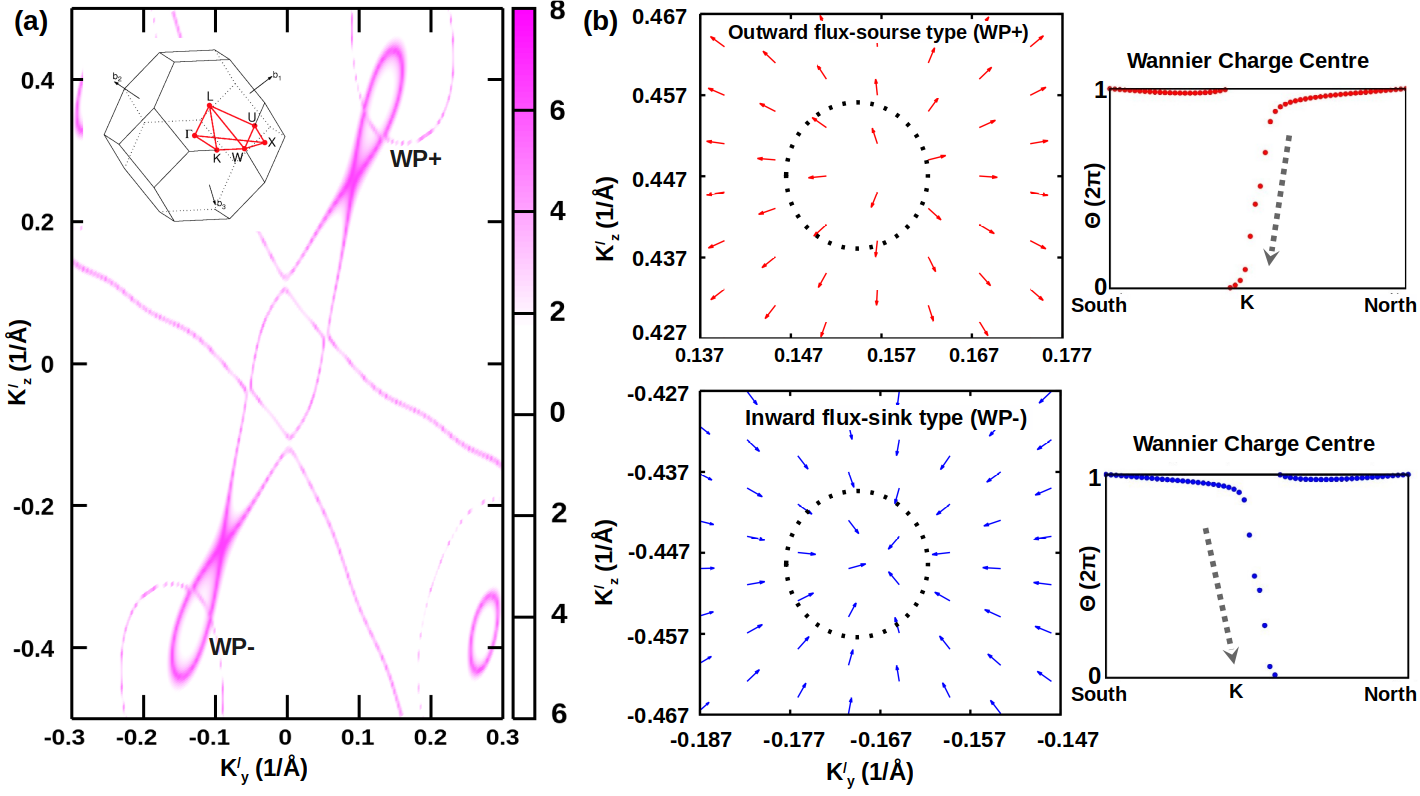}
\caption{Calculated Weyl nodes and Fermi arcs for the are shown for Co$_2$TiSn with chemical potential 278 mev above the Fermi energy. (a) shows the Weyl nodes of two opposite Chern numbers (WP+ and WP-) connected by Fermi arcs in the momentum space in the primitive Brillouin zone (shown in inset)}. (b) Normalized Berry curvatures show source and sink type of flux and the flow chart of the average position of Wannier charge centers (WCC) obtained by the Wilson-loop method applied on a sphere that encloses the two nodes of opposite chirality.
\label{fig:fermi-arcTi}
\end{figure*}

\begin{figure*}[t]
    \includegraphics[width=0.7\linewidth]{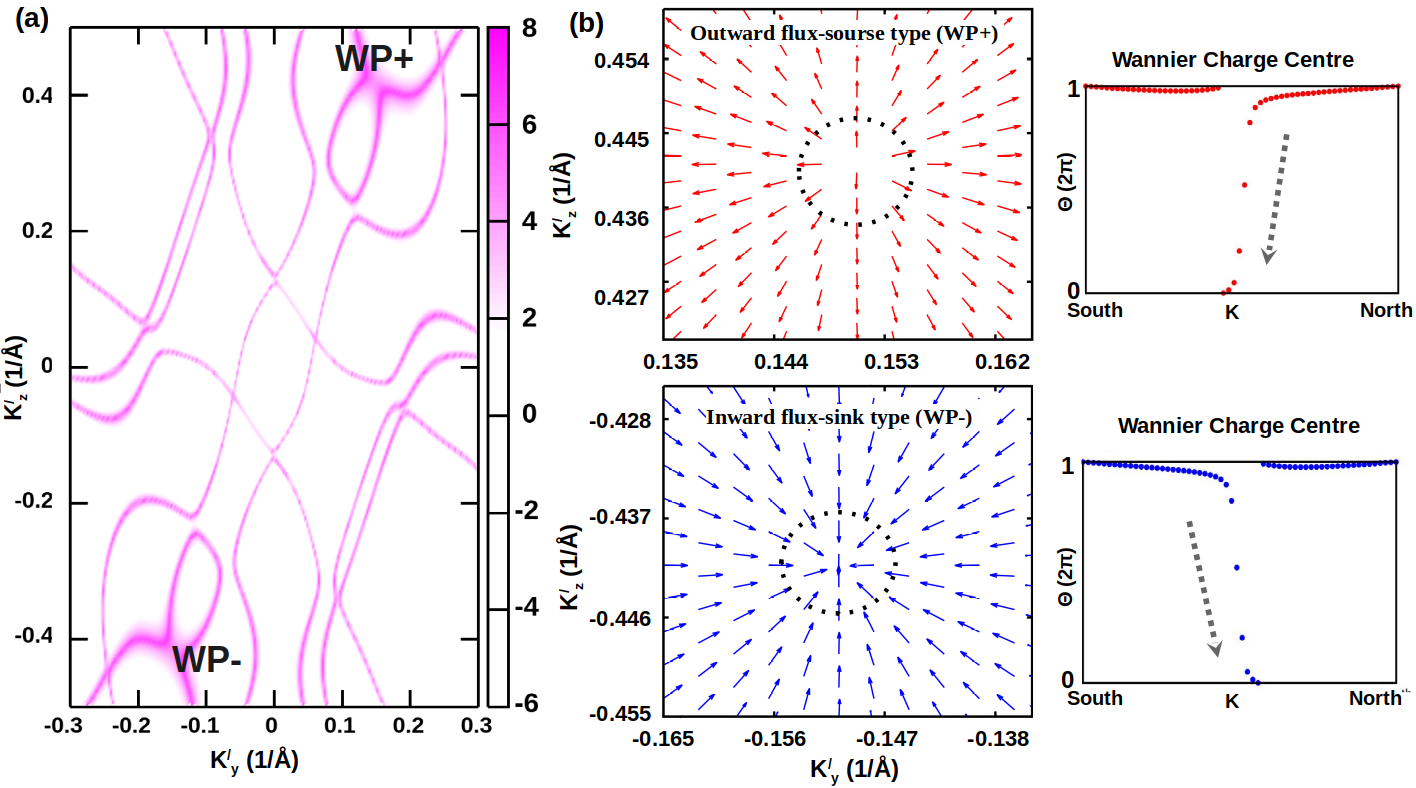}
\caption{Calculated Weyl nodes and Fermi arcs are shown for Co$_2$VSn with chemical potential 227 mev below the Fermi energy. (a) shows the Weyl nodes of two opposite Chern numbers (WP+ and WP-) connected by Fermi arcs in the momentum space in the primitive BZ}. (b) Normalized Berry curvatures show source and sink type of flux  and the flow chart of the average position of Wannier charge centers (WCC) obtained by the Wilson-loop method applied on a sphere that encloses the two nodes of opposite chirality.
\label{fig:fermi-arcV}
\end{figure*}

\begin{figure*}[t]
    \includegraphics[width=0.7\linewidth]{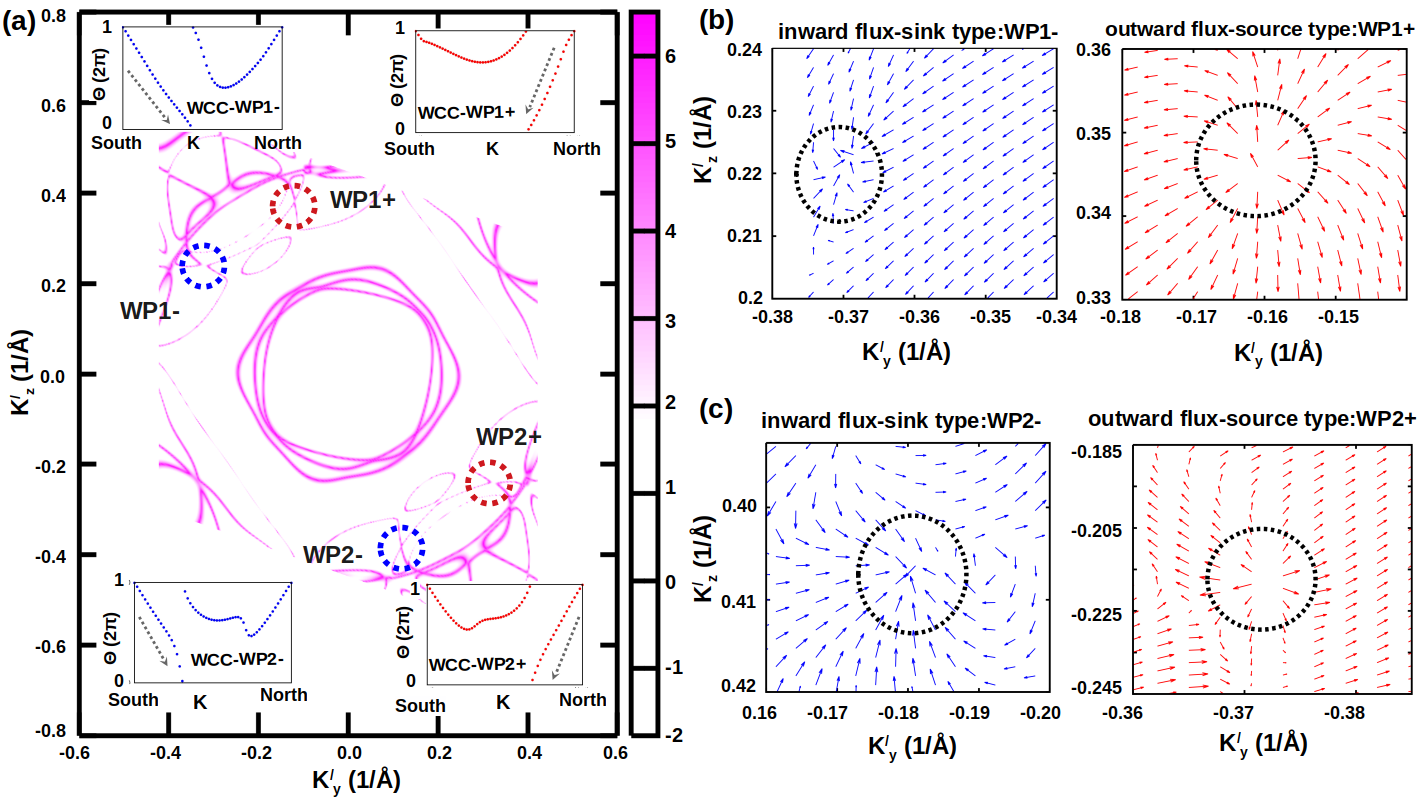}
\caption{Calculated Weyl nodes and Fermi arcs are shown for the Co$_2$Ti$_{0.5}$V$_{0.5}$Sn in the twice supercell in primitive setting}. (a) shows the two equivalent pairs of Weyl nodes with two opposite charality [WP1+ (WP2+) and WP1- (WP2-)] of +1 and -1 respectively. Insets show the WCC for the two paris separately, running in the opposite direction in the K-space.(b) and (c) show  the normalized Berry curvatures for opposite chirality (source and sink type of flux) for WP1 and WP2 respectively.
\label{fig:fermi-arcTiV}
\end{figure*}

We combined the Wannier function basis set based technique starting from pseudo potential plane wave DFT band structure using the energy selective downfolding methodology to extract the Fermi arcs and the normalized Berry curvature. We have used the primitive cell, containing one formula unit of the atoms, for this calculations for the sake of simplicity. The calculated pairs of Weyl nodes with opposite chirality are shown in the table \ref{table:table-2}. To understand the texture of the Weyl points and nodal lines in details, we have plotted these in the Fig.\ref{fig:fermi-arcTi}-\ref{fig:fermi-arcTiV}. The Fig.\ref{fig:fermi-arcTi}(a) shows the Fermi-arc of the Co$_2$TiSn, calculated with chemical potential 278 meV above the Fermi energy. From this figure, it is clearly evident that a pair of Weyl nodes of opposite chirality (+1 and -1) designated by WP+ and WP- respectively, located in the momentum space (-0.274, -0.157, -0.448) and (0.274, 0.157, 0.448) respectively. To clarify the nature of the above mentioned Weyl nodes, we have plotted the normalized Berry curvatures (Fig.\ref{fig:fermi-arcTi}(b)).

The normalized Berry curvatures indicate that the flux at these two Weyl points (WP) are opposite in chiral nature, WP$-$ is sink type, 
where the flux is moving inward whereas, the WP$+$ is more like source type, where the flux comes outward from the point. These 
analysis confirm the opposite chiral nature of these two Weyl nodes. Moreover, we also calculated the flow chart of the average position of Wannier charge centers (WCC) obtained by the Wilson-loop method applied on a sphere that encloses these two nodes of opposite chirality and we found that the WCC is moving in the two 
opposite directions, one is from south to north (WP-) and another is from north to south (WP+), confirming the opposite chiral nature. We did the similar exercise for 
the Co$_2$VSn as shown in Fig.\ref{fig:fermi-arcV}. The Co$_2$VSn Weyl nodes are shown for the chemical potential 227 meV below 
the Fermi energy and we found that the WP of opposite chiralities are located in the (-0.273,-0.150,-0.441) and 
(0.273,0.150,0.441) with Chern numbers -1 and +1 respectively, as shown in the bottom panel of Fig.\ref{fig:fermi-arcV}(b). Point to be noted that, the main features of the Fermi arcs and the normalized Berry curvatures of Co$_2$VSn and Co$_2$TiSn are very similar, which is consistent with the result found from the band structure calculations, which show very similar energy dispersion. The main difference is that the chemical potential for the Fermi arcs, is above and below of the Fermi energy for the Co$_2$TiSn and Co$_2$VSn, respectively.

\begin{table}[htbp]
    \centering
    \resizebox{0.47\textwidth}{!}{%
    	\begin{tabular}{ c | c c c | c | c } 
            \toprule
            \multirow{3}{*}{}
            & \multicolumn{3}{c}{Coordinates} & Chemical & Chern \\
            Compound & \multicolumn{3}{c}{in K-space} & Potential & number \\
            & \multicolumn{3}{c}{} & (meV) & \\
            \midrule
            Co$_2$TiSn                  & -0.270 & -0.167 & -0.448 & 278 & -1 (WP-) \\
                                        &  0.274 &  0.157 &  0.448 & 278 & +1 (WP+)\\
            \midrule
            Co$_2$Ti$_{0.5}$V$_{0.5}$Sn & -0.083 & -0.368 &  0.223 &  7  & -1 (WP1-)  \\
                                        &  0.142 & -0.160 &  0.345 &  18 & +1 (WP1+)\\
                                        &  0.095 &  0.180 & -0.406 &  14 & -1 (WP2-)\\
                                        &  0.035 &  0.371 & -0.217 & -23 & +1 (WP2+)\\
            \midrule                                    
            Co$_2$VSn                   & -0.273 & -0.150 & -0.441 & -227 & -1 (WP-) \\
                                        &  0.273 &  0.150 &  0.441 & -227 & +1 (WP+)\\
            \bottomrule
        \end{tabular}}
    \caption{Representative coordinates in the momentum space in the primitive unit cell setting of the Weyl nodes with their chemical potentials with respect to the Fermi energy together with corresponding Chern number. The Co$_2$TiSn and Co$_2$VSn calculations are done with the primitive cell contaning one fourmula unit of atoms, while the Co$_2$Ti$_{0.5}$V$_{0.5}$Sn calculations are done with the two times supercell of the primitive cell to accommodate $50\%$ doping level.}
    \label{table:table-2}
\end{table}

In the $50\%$ V doped compound, due to the reduction of some symmetry elements due to V substitution within the same crystal structure, the WP pairs do not appear symmetrically in the momentum space, as highlighted in the table \ref{table:table-2}. It should be noted that, due to the two times supercell of the primitive cell is being used to accommodate $50\%$ doping in the DFT band structure calculations followed by Wannier 
interpolation, the number of bands and the Weyl points band crossing near the Fermi energy is also twice compared to that of the undoped compounds. However, if we consider any one pair of the Weyl nodes, say WP1$\pm$, the main features of the Fermi arcs are very similar to that of the two stoichiometric compositions, which ensures the consistency of the calculations. In the Fig.\ref{fig:fermi-arcTiV}(a), we have shown two pairs of Weyl nodes, designated by WP1 and WP2,
and the connected nodal lines between the opposite chiral WP1 (and WP2) points having Chern number +1 and
-1, designated  by WP1+ (WP2+) and WP1- (WP2-) respectively, which are appearing very close to the Fermi
energy. The Fig.\ref{fig:fermi-arcTiV}(b)-(c) show the normalized Berry curvatures of two opposite
chirality Weyl points for the both WP1 and WP2, respectively. Important to note that, for the doped
compound the Weyl points are appearing very close to the Fermi energy, however not exactly at the same
chemical potentials, resulting the Fermi arcs connected through these Weyl points forming some sort of
nodal lines (as evident from the Fig.\ref{fig:fermi-arcTiV}(a)). The chemical potentials of the
corresponding Weyl points were mentioned in the table \ref{table:table-2}. The nature of the two pairs of the Weyl nodes are clear from the sink and source type of normalized Berry curvatures and the oppositely running WCC as shown in the inset of Fig.\ref{fig:fermi-arcTiV}(a). From these analysis we showed that
via chemical doping we can tune the nodal line position in the energy spectrum as well as the nature of the Fermi arcs.

\subsection{Anomalous Hall conductivity}
The anomalous Hall effect/conductivity is direct consequence of the Berry curvature of the electronic band structure near the Fermi level, which act as pseudo-magnetic field in momentum space.The Berry curvature near the Fermi level introduces a transverse  momentum in electron motion and derives  large  anomalous Hall conductivity(AHC). So in our case the middle composition is interest of research due to the fact that Weyl nodes are situated near the Fermi level.To substantiate our compositional dependent theoretical analysis on variation of position of Weyl point in band structure, we theoretically calculated the AHC with the expectation that large AHC should exhibit around the mid composition. To calculate the intrinsic anomalous Hall conductivity (AHC) for the pure and doped systems, the conventional unit cells have been considered (unless otherwise specified). The Bloch wave functions are projected into maximally localized Wannier functions in order to compute the intrinsic AHC. The intrinsic AHC is proportional to the Brillouin zone (BZ) summation of the Berry curvature over all occupied states \cite{pizzi2020wannier90, kubler2012berry,PhysRevB.74.195118}  

\begin{equation}
\sigma^{xy}= -\frac{e^2}{\hbar} \sum_n \int_{BZ} \frac{d\textbf{k}}{(2\pi)^3} \Omega_{n,z}(\textbf{k}) f_n(\textbf{k}),
\label{eq:ahc}
\end{equation}

where $f_n(\textbf{k})$ is the Fermi distribution function for the band $n$, $\Omega_{n,z}(\textbf{k})$ is the $z$ component of 
the Berry curvature at the wave vector $\textbf{k}$. The Berry curvature is related to the Berry connection ($A_n(\textbf{k})$) as 

\begin{equation}
\Omega_n(\textbf{k})= \nabla_\textbf{k} \times A_n(\textbf{k}),
\label{eq:curvature}
\end{equation}

where "$n$" is the band index and $A_n(\textbf{k})$ in terms of cell-periodic Bloch states $\ket{u_{n\textbf{k}}} = e^{-i
\textbf{k.r}}\ket{\psi_{n\textbf{k}}}$ is defined as $ A_n(\textbf{k}) = \expval{i\nabla_\textbf{k}}{u_{n\textbf{k}}} $ 
\cite{pizzi2020wannier90}.

\begin{figure}
    \includegraphics[width=0.8\linewidth]{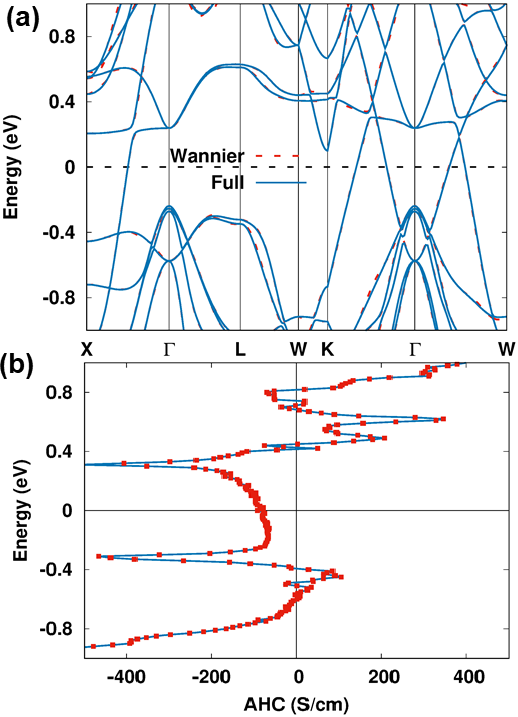}
\caption{ (a) Comparison of Wannier interpolated band structure (red) with the full electronic band structure (blue) of Co$_2$TiSn. 
The Fermi energy is set to 0 $eV$. (b) The calculated intrinsic anomalous Hall conductivity at different energies. The 
conductivity is found to be constant in the vicinity of $E_F$ ($E_F$ $\pm$ 0.2 $eV$).}
\label{fig:figure-6}
\end{figure}

Co$_2$TiSn possesses the fcc $L_{21}$ (space group $\#$ 225) structure which has three mirror planes, $m_x$, $m_y$ and $m_z$ in the absence of any net magnetic moment. These mirror planes protect the gap-less nodal lines in the band structure in $k_x$ = 0, $k_y$ = 0, and $k_z$ = 0 planes\cite{wang2016time, chang2016room}. To compute AHC, the spin-orbit coupling is introduced and the direction of the magnetization has been set along (001). Therefore, due to symmetry breaking of mirror planes the nodal lines in the $k_x$ = 0 and $k_y$ = 0 planes will exhibit a finite band gap while the gap-less nodal line will survive only along the magnetization direction (in the $k_z$ = 0 plane) \cite{wang2016time, chang2016room}. However, due to helical distribution of Berry curvature around this gap-less nodal line, in the mirror plane, the total flux is zero and thereby it does not contribute to the intrinsic AHC \cite{ernst2019anomalous}. Concurrently, the Berry curvature around the broken nodal lines is oriented along the direction of magnetization and contributes to the intrinsic AHC \cite{ernst2019anomalous}. Our calculated intrinsic AHC value (99.38 $S/cm$ and 92.40 $S/cm$ for the primitive cell) at $E_F$ for the pure system is in excellent agreement with the value (100 $S/cm$) reported in literature \cite{ernst2019anomalous}. In addition to this, as it is shown in Fig.~\ref{fig:figure-6} the intrinsic anomalous Hall conductivity is almost constant in the vicinity of $E_F$.
Earlier, we have seen that the band crossing points of the pure compound (Co$_2$TiSn) shift downwards with respect to the Fermi level ($E_F$) as the concentration of $V$ increases (Fig. \ref{fig:figure-2}), and particularly, two of such crossing points which lie in the unoccupied region (for the stoichiometric composition Co$_2$TiSn) come close to $E_F$ as composition approaches toward 50\% V doped Co$_2$TiSn.  Hence, it is expected that the intrinsic AHC will be enhanced for the 50\% V doped Co$_2$TiSn compositions due to strong energy dispersion 
$\sim 2.5$ times higher than the value that obtained near $E_F$ of the pure compound \cite{ernst2019anomalous}.
Considering that the calculation with x= 0.4 doping concentration is computationally very expensive, we have chosen x= 0.5 
to calculate intrinsic AHC, which is expected to provide us maximum value. While simulating AHC as a function of energy, around 0.25 $eV$ above the $E_F$, we get $\sim$ 1.9 times higher AHC than that of the AHC at $E_F$. The same has been further confirmed when the bands are filled along the $+ve$ energy (w.r.t $E_F$), which is achieved in the form of V doping. The calculated AHC along the direction of magnetization at $E_F$ in 50 \% V doped Co$_2$TiSn is found to be 196.84 $S/cm$, nearly twice of the AHC in the pure system(Co$_2$TiSn). Hence, the higher AHC value (in 50 $\%$ doped system as well as at $\sim$ 0.25 $eV$ above the $E_F$ in the pure compound) is attributed to the presence of nodal lines that are very close to $E_F$. 

\par
\section{Conclusion}
To summarise, we performed $\textit{ab-initio}$ calculations on the Co-based Heusler compounds Co$\textsubscript{2}Ti
\textsubscript{1-x}V\textsubscript{x}$Sn with x = 0.0, 0.2, 0.4, 0.6, 0.8 and 1.0. We have calculated the band structures, Bloch spectral functions and DOS using KKR-GF methods. We have also calculated the Fermi arcs, normalized Berry curvatures, WCC and intrinsic AHC for the x = 0.0, 0.5 and 1.0 compositions using Wannier90 interpolation of the plane wave pseudo-potential band structures. We found that nodal lines shift with V substitution and the point of highest energy of the nodal line responsible for Weyl nodes tunes with the Fermi energy for Co\textsubscript{2}Ti\textsubscript{0.6}V\textsubscript{0.4}Sn. For composition between x= 0.6 and 1 the nodel line lie entirely below the Fermi energy. We observed a half-metallic character for the entire range of composition. The magnetic moment on each Co atom as a function of V concentration increases linearly up to x=0.4 and thereafter, it starts decreasing.A detailed investigation reveals that the signature of the Weyl nodes and the Fermi arcs are more prominent near the Fermi energy for the 50$\%$ V doped compound in comparison to the stochiometric compounds, which emphasize the importance of the chemical doping in the present series of compounds. The intrinsic AHC was found to increase by nearly twice for the 50\% doped system as compared to the undoped composition. Our study suggests that Co\textsubscript{2}Ti\textsubscript{1-x}V\textsubscript{x}Sn series of Heusler alloys in general and Co\textsubscript{2}Ti\textsubscript{0.6}V\textsubscript{0.4}Sn composition in particular is important to investigate Weyl physics and various exotic transport phenomena. 
\par
\section{Acknowledgments}
SS thanks Science and Engineering Research Board of India for financial support through the award of Ramanujan Fellowship (grant no: SB/S2/RJN-015/2017), Early Career Research Award (grant no: ECR/2017/003186). SK thanks the Department of Science and Technology (DST), Govt. of India for providing INSPIRE research funding (Grant No.DST/INSPIRE/04/2016/000431; IFA16-MS91). S.W.D and J.M would like to thank CEDAMNF project financed by the Ministry of Education, Youth and Sports of Czech Republic, Project No. CZ.02.1.01/0.0/0.0/15.003/0000358 and also for the support by the GA\v{C}R via the project 20-18725S. KKD and GKS acknowledge the DST-INSPIRE scheme for support through a fellowship. 

Payal Chaudhary and Surasree Sadhukhan contributed
equally to this work.
\par
\medskip
*ssingh.mst@itbhu.ac.in

%
\appendix

\section{Electronic structures of stoichiometric compounds}
\begin{figure}
\center
    \includegraphics[width=1\linewidth]{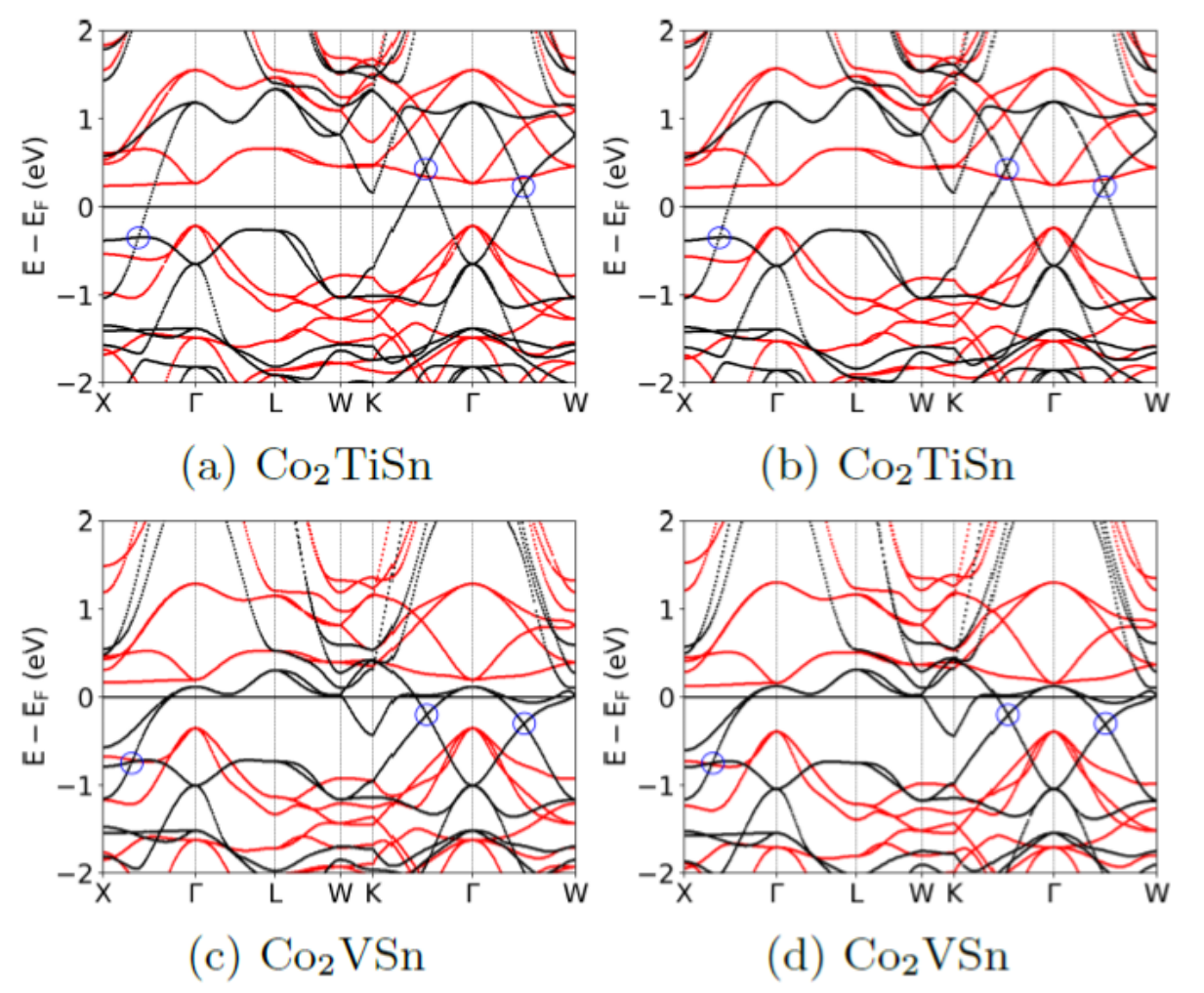}
 \caption{Spin-polarized band structures of Co\textsubscript{2}TiSn and Co\textsubscript{2}VSn with (a),(c) calculated lattice parameters and (b),(d) experimental lattice parameters. The black lines represent majority spin states, and the red lines represent minority spin states. The blue circles show the majority band crossings that form nodal lines in the momentum space.}
\label{fig:figure-7}
\end{figure}

First, we look at the band structure of Co$_2$TiSn (Fig. \ref{fig:figure-7}(a)). When SOC is not considered, the compound possesses three mirror symmetries, $M_x$, $M_y$, and $M_z$ (along the planes $k_x = 0$, $k_y = 0$, and $k_z = 0$), and three $C_{4}$ rotation axes $k_{x}$, $k_{y}$, and $k_{z}$. Due to these symmetries, there exist nodal lines in the momentum space on each of the three planes, formed by the crossings of the spin-up states of the Co-Ti hybridized 3$\textit{d}$ electron bands. These crossings can be seen in the band structures in Fig. \ref{fig:figure-7} along the $\Gamma-X$, $\Gamma-K$ and $\Gamma-W$ high-symmetry lines, circled in blue. The energy of these nodal lines oscillates around the Fermi energy, reaching a maximum energy of 0.45 eV, along the $\Gamma-K$ direction, and a minimum of -0.35 eV, along the $\Gamma-X$ direction. The high-symmetry points and the directions followed in the band structures are the same as illustrated in Fig. \ref{fig:figure-2}(b).

When Ti is replaced by V, from Co$_2$TiSn to Co$_2$VSn, an additional 3\textit{d} electron is added. As a result, in Co$_2$VSn, one of the spin-up Co-V (previously Co-Ti) hybridized 3\textit{d} bands, which had higher energy in Co$_2$TiSn, now lies on the Fermi energy. The Fermi energy itself increases, and the other states below the Fermi level only shift rigidly maintaining the same band shape. There is no change in the shape of the spin-down states across all energy levels, and the half-metallic character is retained.

The minority spin band gap and total magnetic moment per formula unit are 0.491 eV and 2.03 $\mu_{B}$, and 0.546 eV and 3.01 $\mu_{B}$, for Co$_2$TiSn and Co$_2$VSn respectively, in good agreement with literature \cite{doi:10.1063/1.1853899,hickey2006Fermi,kandpal2007calculated,barth2010itinerant,aguayo2011density}. The difference in the minority spin band gap, which arises from the splitting of the anti-bonding Co 3\textit{d} orbitals \cite{galanakis2002slater,aguayo2011density}, is nearly 0.05 eV. This difference may be attributed to the difference in lattice parameters of the two compounds \cite{kandpal2007calculated}.

Fig. \ref{fig:figure-7} (a) and (c) show the spin-polarised, non spin-orbit coupled band structures of Co$_2$TiSn and Co$_2$VSn obtained using the calculated lattice parameters. Non-SOC band structures provide an insight into the symmetry of the bands, and how the symmetry changes when SOC is considered and Weyl nodes are formed. Through these structures we can also study how the hybridisation of these bands changes on inclusion of spin-orbit coupling. When SOC is considered, the mirror symmetries are broken according to the direction of magnetization. Due to this, the nodal lines gap out results in Weyl points, and there is a large Berry curvature in the vicinity of Fermi level.
\begin{figure}
\center
\includegraphics[scale = 0.5]{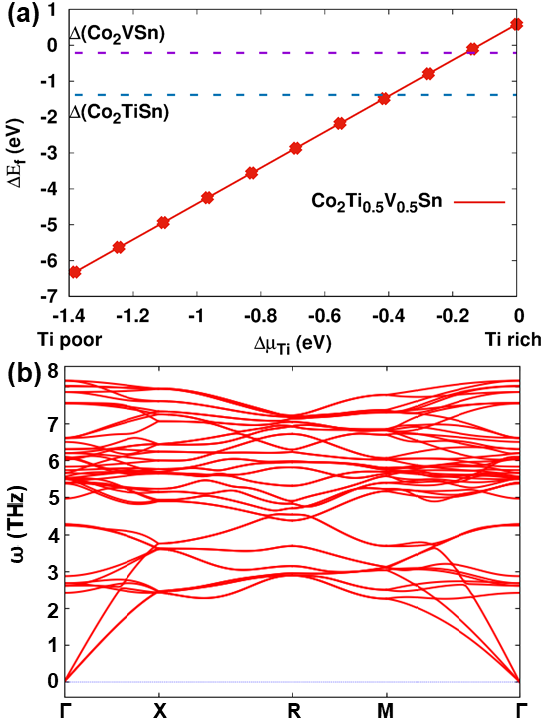}
\caption{ (a) Defect formation energy of $Co_2Ti_{0.5}V_{0.5}Sn$ varies with the chemical potential of $Ti$. The enthalpy of formation for two end-compounds, $Co_2TiSn$ and $Co_2VSn$, are also shown in the same plot. (b) Phonon dispersion relations of $Co_2Ti_{0.5}V_{0.5}Sn$ crystal along the high-symmetry lines of the optimized structure.}
\label{fig:cts-apndx}
\end{figure}
\begin{figure}
\center
\includegraphics[scale = 0.5]{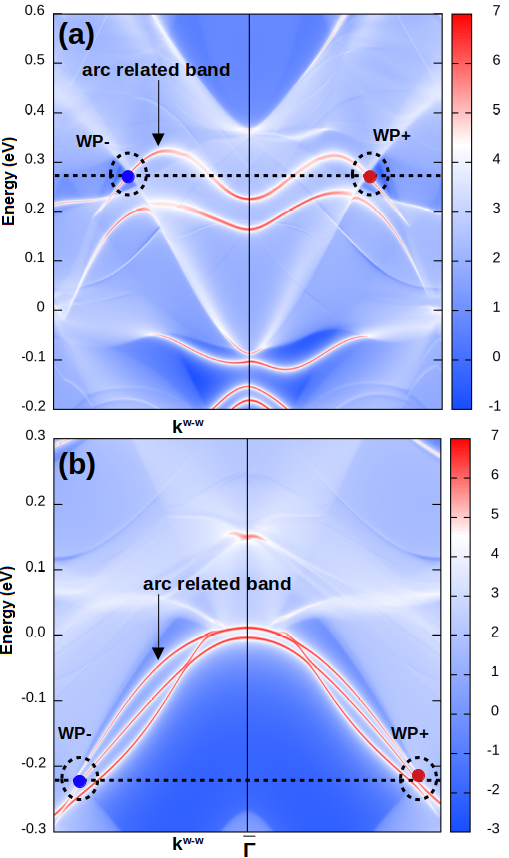}
\caption{Calculated surface states using slab calculations for the stoichiometric compositions (a) Co$_2$TiSn and (b) Co$_2$VSn for the [100] surface. The surface spectral function crossing one pair of Weyl points of opposite chirality in the BZ for Co$_2$TiSn and Co$_2$VSn respectively. The dashed line represents the chemical potentials where the WP's are appearing. The spectral function responsible for showing Fermi arcs are marked in the figure. Fermi energy set at zero in the energy axis. }
\label{fig:SS}
\end{figure}

\section{Thermodynamic and dynamical stability}

In order to examine the stability of the pure and doped compounds we have calculated the enthalpy of formation ($\Delta$) for two end-compounds, $Co_2TiSn$ and $Co_2VSn$, as shown in the following plot (Fig.~\ref{fig:cts-apndx}(a)). The negative values (dotted curves) of $\Delta$ for both the compounds suggest their thermodynamical stability, the same are already synthesized experimentally. Further, we have calculated the defect formation energy ($\Delta E_f$) at the mid composition ($Co_2Ti_{0.5}V_{0.5}Sn$), with respect to $Co_2TiSn$, using the relation

\begin{equation}
\begin{aligned}
\Delta E_f =
E(Co_2Ti_{0.5}V_{0.5}Sn) - E(Co_2TiSn)- \sum_i \Delta n_i \mu_{Ti},
\end{aligned}
\label{eq:dfe}
\end{equation}

where the first two terms are the total energies of doped and pure systems respectively. $\Delta n_i$ is the change in the number of atoms of species $i$, +ve for V (introduced) and -ve for Ti (removed). $\mu_{Ti}$ is the chemical potential of the element $Ti$. Since we are calculating the $\Delta E_f$ with respect to $Co_2TiSn$, the range of elemental chemical potential of the species $Ti$ is defined as $\mu_{Ti} (bulk)+ \Delta (Co_2TiSn) \leq \mu_{Ti} \leq \mu_{Ti} (bulk)$. The variation of $\Delta E_f$, versus $\mu_{Ti}$, in case of $Co_2Ti_{0.5}V_{0.5}Sn$ is shown in the same figure. The figure suggests that $\Delta E_f$ is more -ve when we approach towards the $Ti$ poor condition which also suggests the possibility of vacancy formation. The results for $Co_2Ti_{0.5}V_{0.5}Sn$ further indicates that the compound will certainly be thermodynamically stable at any doping level. Because, the thermodynamically stable 50\% doping composition is the highest doping concentration in $Co_2Ti_{1-x}V_xSn$ series of Heusler alloys in which two end-compounds are also stable. The stability of $Co_2Ti_{0.5}V_{0.5}Sn$ is further examined by calculating its phonon dispersion curve along the appropriate high-symmetry paths of the optimized structure. The absence of imaginary frequency in Fig.~\ref{fig:cts-apndx}(b) further suggests the dynamical stability of $Co_2Ti_{1-x}V_xSn$ series of Heusler compounds. 

\section{Surface spectral functions}
The calculated surface spectral function of [100] surface using slab model calculations using the Wannier derived Hamiltonian for the stoichiometric compositions Co$_2$TiSn and Co$_2$VSn are shown in the Fig.~\ref{fig:SS}(a) and (b) respectively. To understand the Fermi arcs in the surface state, we have plotted the surface spectral function along the k-path (k$^{ww}$) connecting the pair of Weyl points of opposite chirality, say, WP+ and WP-. The figure clearly shows the bands responsible for the Fermi arcs connecting the WP- and WP+ points appearing in the surface state as well.


\begin{thebibliography}{74}%
\makeatletter
\providecommand \@ifxundefined [1]{%
 \@ifx{#1\undefined}
}%
\providecommand \@ifnum [1]{%
 \ifnum #1\expandafter \@firstoftwo
 \else \expandafter \@secondoftwo
 \fi
}%
\providecommand \@ifx [1]{%
 \ifx #1\expandafter \@firstoftwo
 \else \expandafter \@secondoftwo
 \fi
}%
\providecommand \natexlab [1]{#1}%
\providecommand \enquote  [1]{``#1''}%
\providecommand \bibnamefont  [1]{#1}%
\providecommand \bibfnamefont [1]{#1}%
\providecommand \citenamefont [1]{#1}%
\providecommand \href@noop [0]{\@secondoftwo}%
\providecommand \href [0]{\begingroup \@sanitize@url \@href}%
\providecommand \@href[1]{\@@startlink{#1}\@@href}%
\providecommand \@@href[1]{\endgroup#1\@@endlink}%
\providecommand \@sanitize@url [0]{\catcode `\\12\catcode `\$12\catcode
  `\&12\catcode `\#12\catcode `\^12\catcode `\_12\catcode `\%12\relax}%
\providecommand \@@startlink[1]{}%
\providecommand \@@endlink[0]{}%
\providecommand \url  [0]{\begingroup\@sanitize@url \@url }%
\providecommand \@url [1]{\endgroup\@href {#1}{\urlprefix }}%
\providecommand \urlprefix  [0]{URL }%
\providecommand \Eprint [0]{\href }%
\providecommand \doibase [0]{https://doi.org/}%
\providecommand \selectlanguage [0]{\@gobble}%
\providecommand \bibinfo  [0]{\@secondoftwo}%
\providecommand \bibfield  [0]{\@secondoftwo}%
\providecommand \translation [1]{[#1]}%
\providecommand \BibitemOpen [0]{}%
\providecommand \bibitemStop [0]{}%
\providecommand \bibitemNoStop [0]{.\EOS\space}%
\providecommand \EOS [0]{\spacefactor3000\relax}%
\providecommand \BibitemShut  [1]{\csname bibitem#1\endcsname}%
\let\auto@bib@innerbib\@empty
\bibitem [{\citenamefont {Hu}\ \emph {et~al.}(2019{\natexlab{a}})\citenamefont
  {Hu}, \citenamefont {Xu}, \citenamefont {Ni},\ and\ \citenamefont
  {Mao}}]{hu2019transport}%
  \BibitemOpen
  \bibfield  {author} {\bibinfo {author} {\bibfnamefont {J.}~\bibnamefont
  {Hu}}, \bibinfo {author} {\bibfnamefont {S.-Y.}\ \bibnamefont {Xu}}, \bibinfo
  {author} {\bibfnamefont {N.}~\bibnamefont {Ni}},\ and\ \bibinfo {author}
  {\bibfnamefont {Z.}~\bibnamefont {Mao}},\ }\bibfield  {title} {\bibinfo
  {title} {Transport of topological semimetals},\ }\href@noop {} {\bibfield
  {journal} {\bibinfo  {journal} {Annu. Rev. Mater. Res.}\
  }\textbf {\bibinfo {volume} {49}},\ \bibinfo {pages} {207} (\bibinfo {year}
  {2019}{\natexlab{a}})}\BibitemShut {NoStop}%
\bibitem [{\citenamefont {Armitage}\ \emph {et~al.}(2018)\citenamefont
  {Armitage}, \citenamefont {Mele},\ and\ \citenamefont
  {Vishwanath}}]{RevModPhys.90.015001}%
  \BibitemOpen
  \bibfield  {author} {\bibinfo {author} {\bibfnamefont {N.~P.}\ \bibnamefont
  {Armitage}}, \bibinfo {author} {\bibfnamefont {E.~J.}\ \bibnamefont {Mele}},\
  and\ \bibinfo {author} {\bibfnamefont {A.}~\bibnamefont {Vishwanath}},\
  }\bibfield  {title} {\bibinfo {title} {Weyl and Dirac semimetals in
  three-dimensional solids},\ }\href
  {https://doi.org/10.1103/RevModPhys.90.015001} {\bibfield  {journal}
  {\bibinfo  {journal} {Rev. Mod. Phys.}\ }\textbf {\bibinfo {volume} {90}},\
  \bibinfo {pages} {015001} (\bibinfo {year} {2018})}\BibitemShut {NoStop}%
\bibitem [{\citenamefont {Shekhar}\ \emph {et~al.}(2015)\citenamefont
  {Shekhar}, \citenamefont {Nayak}, \citenamefont {Sun}, \citenamefont
  {Schmidt}, \citenamefont {Nicklas}, \citenamefont {Leermakers}, \citenamefont
  {Zeitler}, \citenamefont {Skourski}, \citenamefont {Wosnitza}, \citenamefont
  {Liu} \emph {et~al.}}]{shekhar2015extremely}%
  \BibitemOpen
  \bibfield  {author} {\bibinfo {author} {\bibfnamefont {C.}~\bibnamefont
  {Shekhar}}, \bibinfo {author} {\bibfnamefont {A.~K.}\ \bibnamefont {Nayak}},
  \bibinfo {author} {\bibfnamefont {Y.}~\bibnamefont {Sun}}, \bibinfo {author}
  {\bibfnamefont {M.}~\bibnamefont {Schmidt}}, \bibinfo {author} {\bibfnamefont
  {M.}~\bibnamefont {Nicklas}}, \bibinfo {author} {\bibfnamefont
  {I.}~\bibnamefont {Leermakers}}, \bibinfo {author} {\bibfnamefont
  {U.}~\bibnamefont {Zeitler}}, \bibinfo {author} {\bibfnamefont
  {Y.}~\bibnamefont {Skourski}}, \bibinfo {author} {\bibfnamefont
  {J.}~\bibnamefont {Wosnitza}}, \bibinfo {author} 
  {\bibfnamefont
  {Z.}~\bibnamefont {Liu}}, 
  \bibinfo {author} 
  {\bibfnamefont
  {Y.}~\bibnamefont {Chen}}, \bibinfo {author} 
  {\bibfnamefont
  {W.}~\bibnamefont {Schnelle}}, \bibinfo {author} 
  {\bibfnamefont
  {H.}~\bibnamefont {Borrmann}}, \bibinfo {author} 
  {\bibfnamefont
  {Y.}~\bibnamefont {Grin}}, \bibinfo {author} 
  {\bibfnamefont
  {C.}~\bibnamefont {Felser}},\ and\ \bibinfo {author} 
  {\bibfnamefont
  {B.}~\bibnamefont {Yan}},\ }\bibfield  {title} {\bibinfo
  {title} {Extremely large magnetoresistance and ultrahigh mobility in the
  topological Weyl semimetal candidate NbP},\ }\href@noop {} {\bibfield
  {journal} {\bibinfo  {journal} {Nat. Phys.}\ }\textbf {\bibinfo {volume}
  {11}},\ \bibinfo {pages} {645} (\bibinfo {year} {2015})}\BibitemShut
  {NoStop}%
\bibitem [{\citenamefont {Wan}\ \emph {et~al.}(2011)\citenamefont {Wan},
  \citenamefont {Turner}, \citenamefont {Vishwanath},\ and\ \citenamefont
  {Savrasov}}]{wan2011topological}%
  \BibitemOpen
  \bibfield  {author} {\bibinfo {author} {\bibfnamefont {X.}~\bibnamefont
  {Wan}}, \bibinfo {author} {\bibfnamefont {A.~M.}\ \bibnamefont {Turner}},
  \bibinfo {author} {\bibfnamefont {A.}~\bibnamefont {Vishwanath}},\ and\
  \bibinfo {author} {\bibfnamefont {S.~Y.}\ \bibnamefont {Savrasov}},\
  }\bibfield  {title} {\bibinfo {title} {Topological semimetal and Fermi-arc
  surface states in the electronic structure of pyrochlore iridates},\
  }\href@noop {} {\bibfield  {journal} {\bibinfo  {journal} {Phys. Rev. B}\ }\textbf {\bibinfo {volume} {83}},\ \bibinfo {pages} {205101} (\bibinfo
  {year} {2011})}\BibitemShut {NoStop}%
\bibitem [{\citenamefont {Jia}\ \emph {et~al.}(2016)\citenamefont {Jia},
  \citenamefont {Xu},\ and\ \citenamefont {Hasan}}]{jia2016weyl}%
  \BibitemOpen
  \bibfield  {author} {\bibinfo {author} {\bibfnamefont {S.}~\bibnamefont
  {Jia}}, \bibinfo {author} {\bibfnamefont {S.-Y.}\ \bibnamefont {Xu}},\ and\
  \bibinfo {author} {\bibfnamefont {M.}\ \bibnamefont {Zahid Hasan}},\ }\bibfield
  {title} {\bibinfo {title} {Weyl semimetals, Fermi arcs and chiral
  anomalies},\ }\href@noop {} {\bibfield  {journal} {\bibinfo  {journal}
  {Nat. Mater.}\ }\textbf {\bibinfo {volume} {15}},\ \bibinfo {pages}
  {1140} (\bibinfo {year} {2016})}\BibitemShut {NoStop}%
\bibitem [{\citenamefont {Son}\ and\ \citenamefont
  {Spivak}(2013)}]{son2013chiral}%
  \BibitemOpen
  \bibfield  {author} {\bibinfo {author} {\bibfnamefont {D.}~\bibnamefont
  {Son}}\ and\ \bibinfo {author} {\bibfnamefont {B.}~\bibnamefont {Spivak}},\
  }\bibfield  {title} {\bibinfo {title} {Chiral anomaly and classical negative
  magnetoresistance of Weyl metals},\ }\href@noop {} {\bibfield  {journal}
  {\bibinfo  {journal} {Phys. Rev. B}\ }\textbf {\bibinfo {volume} {88}},\
  \bibinfo {pages} {104412} (\bibinfo {year} {2013})}\BibitemShut {NoStop}%
\bibitem [{\citenamefont {Burkov}(2014)}]{burkov2014anomalous}%
  \BibitemOpen
  \bibfield  {author} {\bibinfo {author} {\bibfnamefont {A.~A.}~\bibnamefont
  {Burkov}},\ }\bibfield  {title} {\bibinfo {title} {Anomalous Hall effect in
  Weyl metals},\ }\href@noop {} {\bibfield  {journal} {\bibinfo  {journal}
  {Phys. Rev. Lett.}\ }\textbf {\bibinfo {volume} {113}},\ \bibinfo
  {pages} {187202} (\bibinfo {year} {2014})}\BibitemShut {NoStop}%
\bibitem [{\citenamefont {Shekhar}\ \emph {et~al.}(2018)\citenamefont
  {Shekhar}, \citenamefont {Kumar}, \citenamefont {Grinenko}, \citenamefont
  {Singh}, \citenamefont {Sarkar}, \citenamefont {Luetkens}, \citenamefont
  {Wu}, \citenamefont {Zhang}, \citenamefont {Komarek}, \citenamefont
  {Kampert}, \citenamefont {Skourski}, \citenamefont {Wosnitza}, \citenamefont
  {Schnelle}, \citenamefont {McCollam}, \citenamefont {Zeitler}, \citenamefont
  {K{\"u}bler}, \citenamefont {Yan}, \citenamefont {Klauss}, \citenamefont
  {Parkin},\ and\ \citenamefont {Felser}}]{Shekhar9140}%
  \BibitemOpen
  \bibfield  {author} {\bibinfo {author} {\bibfnamefont {C.}~\bibnamefont
  {Shekhar}}, \bibinfo {author} {\bibfnamefont {N.}~\bibnamefont {Kumar}},
  \bibinfo {author} {\bibfnamefont {V.}~\bibnamefont {Grinenko}}, \bibinfo
  {author} {\bibfnamefont {S.}~\bibnamefont {Singh}}, \bibinfo {author}
  {\bibfnamefont {R.}~\bibnamefont {Sarkar}}, \bibinfo {author} {\bibfnamefont
  {H.}~\bibnamefont {Luetkens}}, \bibinfo {author} {\bibfnamefont {S.-C.}\
  \bibnamefont {Wu}}, \bibinfo {author} {\bibfnamefont {Y.}~\bibnamefont
  {Zhang}}, \bibinfo {author} {\bibfnamefont {A.~C.}\ \bibnamefont {Komarek}},
  \bibinfo {author} {\bibfnamefont {E.}~\bibnamefont {Kampert}}, \bibinfo
  {author} {\bibfnamefont {Y.}~\bibnamefont {Skourski}}, \bibinfo {author}
  {\bibfnamefont {J.}~\bibnamefont {Wosnitza}}, \bibinfo {author}
  {\bibfnamefont {W.}~\bibnamefont {Schnelle}}, \bibinfo {author}
  {\bibfnamefont {A.}~\bibnamefont {McCollam}}, \bibinfo {author}
  {\bibfnamefont {U.}~\bibnamefont {Zeitler}}, \bibinfo {author} {\bibfnamefont
  {J.}~\bibnamefont {K{\"u}bler}}, \bibinfo {author} {\bibfnamefont
  {B.}~\bibnamefont {Yan}}, \bibinfo {author} {\bibfnamefont {H.-H.}\
  \bibnamefont {Klauss}}, \bibinfo {author} {\bibfnamefont {S.~S.~P.}\
  \bibnamefont {Parkin}},\ and\ \bibinfo {author} {\bibfnamefont
  {C.}~\bibnamefont {Felser}},\ }\bibfield  {title} {\bibinfo {title}
  {Anomalous Hall effect in Weyl semimetal half-Heusler compounds RPtBi (R = Gd
  and Nd)},\ }\href {https://doi.org/10.1073/pnas.1810842115} {\bibfield
  {journal} {\bibinfo  {journal} {Proc. Natl. Acad. Sci. USA}\ }\textbf
  {\bibinfo {volume} {115}},\ \bibinfo {pages} {9140} (\bibinfo {year}
  {2018})}\BibitemShut {NoStop}%
\bibitem [{\citenamefont {Ikhlas}\ \emph {et~al.}(2017)\citenamefont {Ikhlas},
  \citenamefont {Tomita}, \citenamefont {Koretsune}, \citenamefont {Suzuki},
  \citenamefont {Nishio-Hamane}, \citenamefont {Arita}, \citenamefont {Otani},\
  and\ \citenamefont {Nakatsuji}}]{ikhlas2017large}%
  \BibitemOpen
  \bibfield  {author} {\bibinfo {author} {\bibfnamefont {M.}~\bibnamefont
  {Ikhlas}}, \bibinfo {author} {\bibfnamefont {T.}~\bibnamefont {Tomita}},
  \bibinfo {author} {\bibfnamefont {T.}~\bibnamefont {Koretsune}}, \bibinfo
  {author} {\bibfnamefont {M.-T.}\ \bibnamefont {Suzuki}}, \bibinfo {author}
  {\bibfnamefont {D.}~\bibnamefont {Nishio-Hamane}}, \bibinfo {author}
  {\bibfnamefont {R.}~\bibnamefont {Arita}}, \bibinfo {author} {\bibfnamefont
  {Y.}~\bibnamefont {Otani}},\ and\ \bibinfo {author} {\bibfnamefont
  {S.}~\bibnamefont {Nakatsuji}},\ }\bibfield  {title} {\bibinfo {title} {Large
  anomalous Nernst effect at room temperature in a chiral antiferromagnet},\
  }\href@noop {} {\bibfield  {journal} {\bibinfo  {journal} {Nat. Phys.}\
  }\textbf {\bibinfo {volume} {13}},\ \bibinfo {pages} {1085} (\bibinfo {year}
  {2017})}\BibitemShut {NoStop}%
\bibitem [{\citenamefont {Sakai}\ \emph {et~al.}(2018)\citenamefont {Sakai},
  \citenamefont {Mizuta}, \citenamefont {Nugroho}, \citenamefont {Sihombing},
  \citenamefont {Koretsune}, \citenamefont {Suzuki}, \citenamefont {Takemori},
  \citenamefont {Ishii}, \citenamefont {Nishio-Hamane}, \citenamefont {Arita}
  \emph {et~al.}}]{sakai2018giant}%
  \BibitemOpen
  \bibfield  {author} {\bibinfo {author} {\bibfnamefont {A.}~\bibnamefont
  {Sakai}}, \bibinfo {author} {\bibfnamefont {Y.~P.}\ \bibnamefont {Mizuta}},
  \bibinfo {author} {\bibfnamefont {A.~A.}\ \bibnamefont {Nugroho}}, \bibinfo
  {author} {\bibfnamefont {R.}~\bibnamefont {Sihombing}}, \bibinfo {author}
  {\bibfnamefont {T.}~\bibnamefont {Koretsune}}, \bibinfo {author}
  {\bibfnamefont {M.-T.}\ \bibnamefont {Suzuki}}, \bibinfo {author}
  {\bibfnamefont {N.}~\bibnamefont {Takemori}}, \bibinfo {author}
  {\bibfnamefont {R.}~\bibnamefont {Ishii}}, \bibinfo {author} {\bibfnamefont
  {D.}~\bibnamefont {Nishio-Hamane}}, \bibinfo {author} {\bibfnamefont
  {R.}~\bibnamefont {Arita}}, \bibinfo {author} {\bibfnamefont
  {P.}~\bibnamefont {Goswami}},\ and\ \bibinfo {author} {\bibfnamefont
  {S.}~\bibnamefont {Nakatsuji}},\ }\bibfield  {title} {\bibinfo
  {title} {Giant anomalous Nernst effect and quantum-critical scaling in a
  ferromagnetic semimetal},\ }\href@noop {} {\bibfield  {journal} {\bibinfo
  {journal} {Nat. Phys.}\ }\textbf {\bibinfo {volume} {14}},\ \bibinfo
  {pages} {1119} (\bibinfo {year} {2018})}\BibitemShut {NoStop}%
\bibitem [{\citenamefont {Taguchi}\ \emph {et~al.}(2016)\citenamefont
  {Taguchi}, \citenamefont {Imaeda}, \citenamefont {Sato},\ and\ \citenamefont
  {Tanaka}}]{taguchi2016photovoltaic}%
  \BibitemOpen
  \bibfield  {author} {\bibinfo {author} {\bibfnamefont {K.}~\bibnamefont
  {Taguchi}}, \bibinfo {author} {\bibfnamefont {T.}~\bibnamefont {Imaeda}},
  \bibinfo {author} {\bibfnamefont {M.}~\bibnamefont {Sato}},\ and\ \bibinfo
  {author} {\bibfnamefont {Y.}~\bibnamefont {Tanaka}},\ }\bibfield  {title}
  {\bibinfo {title} {Photovoltaic chiral magnetic effect in Weyl semimetals},\
  }\href@noop {} {\bibfield  {journal} {\bibinfo  {journal} {Phys. Rev. B}\ }\textbf {\bibinfo {volume} {93}},\ \bibinfo {pages} {201202} (\bibinfo
  {year} {2016})}\BibitemShut {NoStop}%
\bibitem [{\citenamefont {Chan}\ \emph {et~al.}(2017)\citenamefont {Chan},
  \citenamefont {Lindner}, \citenamefont {Refael},\ and\ \citenamefont
  {Lee}}]{chan2017photocurrents}%
  \BibitemOpen
  \bibfield  {author} {\bibinfo {author} {\bibfnamefont {C.-K.}\ \bibnamefont
  {Chan}}, \bibinfo {author} {\bibfnamefont {N.~H.}\ \bibnamefont {Lindner}},
  \bibinfo {author} {\bibfnamefont {G.}~\bibnamefont {Refael}},\ and\ \bibinfo
  {author} {\bibfnamefont {P.~A.}\ \bibnamefont {Lee}},\ }\bibfield  {title}
  {\bibinfo {title} {Photocurrents in Weyl semimetals},\ }\href@noop {}
  {\bibfield  {journal} {\bibinfo  {journal} {Phys. Rev. B}\ }\textbf
  {\bibinfo {volume} {95}},\ \bibinfo {pages} {041104} (\bibinfo {year}
  {2017})}\BibitemShut {NoStop}%
\bibitem [{\citenamefont {de~Juan}\ \emph {et~al.}(2017)\citenamefont
  {de~Juan}, \citenamefont {Grushin}, \citenamefont {Morimoto},\ and\
  \citenamefont {Moore}}]{de2017quantized}%
  \BibitemOpen
  \bibfield  {author} {\bibinfo {author} {\bibfnamefont {F.}~\bibnamefont
  {de~Juan}}, \bibinfo {author} {\bibfnamefont {A.~G.}\ \bibnamefont
  {Grushin}}, \bibinfo {author} {\bibfnamefont {T.}~\bibnamefont {Morimoto}},\
  and\ \bibinfo {author} {\bibfnamefont {J.~E.}\ \bibnamefont {Moore}},\
  }\bibfield  {title} {\bibinfo {title} {Quantized circular photogalvanic
  effect in Weyl semimetals},\ }\href@noop {} {\bibfield  {journal} {\bibinfo
  {journal} {Nat. Commun.}\ }\textbf {\bibinfo {volume} {8}},\
  \bibinfo {pages} {15995} (\bibinfo {year} {2017})}\BibitemShut {NoStop}%
\bibitem [{\citenamefont {Osterhoudt}\ \emph {et~al.}(2019)\citenamefont
  {Osterhoudt}, \citenamefont {Diebel}, \citenamefont {Yang}, \citenamefont
  {Stanco}, \citenamefont {Huang}, \citenamefont {Shen}, \citenamefont {Ni},
  \citenamefont {Moll}, \citenamefont {Ran},\ and\ \citenamefont
  {Burch}}]{osterhoudt2017colossal}%
  \BibitemOpen
  \bibfield  {author} {\bibinfo {author} {\bibfnamefont {G.~B.}\ \bibnamefont
  {Osterhoudt}}, \bibinfo {author} {\bibfnamefont {L.~K.}\ \bibnamefont
  {Diebel}}, \bibinfo {author} {\bibfnamefont {M. J.}~\bibnamefont {Gray}}, \bibinfo {author} {\bibfnamefont {X.}~\bibnamefont {Yang}},
  \bibinfo {author} {\bibfnamefont {J.}~\bibnamefont {Stanco}}, \bibinfo
  {author} {\bibfnamefont {X.}~\bibnamefont {Huang}}, \bibinfo {author}
  {\bibfnamefont {B.}~\bibnamefont {Shen}}, \bibinfo {author} {\bibfnamefont
  {N.}~\bibnamefont {Ni}}, \bibinfo {author} {\bibfnamefont {P.}~\bibnamefont
  {Moll}}, \bibinfo {author} {\bibfnamefont {Y.}~\bibnamefont {Ran}},\ and\
  \bibinfo {author} {\bibfnamefont {K.~S.}\ \bibnamefont {Burch}},\ }\bibfield
  {title} {\bibinfo {title} {Colossal mid-infrared bulk photovoltaic effect in a type-I Weyl semimetal},\ }\href@noop {} {\bibfield
  {journal} {\bibinfo  {journal} {Nat. Mater.}\ } \textbf {\bibinfo {volume} {18}},\ \bibinfo {pages} {471} (\bibinfo {year} {2019})}\BibitemShut {NoStop}%
\bibitem [{\citenamefont {Morimoto}\ and\ \citenamefont
  {Nagaosa}(2016)}]{morimoto2016topological}%
  \BibitemOpen
  \bibfield  {author} {\bibinfo {author} {\bibfnamefont {T.}~\bibnamefont
  {Morimoto}}\ and\ \bibinfo {author} {\bibfnamefont {N.}~\bibnamefont
  {Nagaosa}},\ }\bibfield  {title} {\bibinfo {title} {Topological nature of
  nonlinear optical effects in solids},\ }\href@noop {} {\bibfield  {journal}
  {\bibinfo  {journal} {Sci. Adv.}\ }\textbf {\bibinfo {volume} {2}},\
  \bibinfo {pages} {e1501524} (\bibinfo {year} {2016})}\BibitemShut {NoStop}%
\bibitem [{\citenamefont {Wu}\ \emph {et~al.}(2017)\citenamefont {Wu},
  \citenamefont {Patankar}, \citenamefont {Morimoto}, \citenamefont {Nair},
  \citenamefont {Thewalt}, \citenamefont {Little}, \citenamefont {Analytis},
  \citenamefont {Moore},\ and\ \citenamefont {Orenstein}}]{wu2017giant}%
  \BibitemOpen
  \bibfield  {author} {\bibinfo {author} {\bibfnamefont {L.}~\bibnamefont
  {Wu}}, \bibinfo {author} {\bibfnamefont {S.}~\bibnamefont {Patankar}},
  \bibinfo {author} {\bibfnamefont {T.}~\bibnamefont {Morimoto}}, \bibinfo
  {author} {\bibfnamefont {N.~L.}\ \bibnamefont {Nair}}, \bibinfo {author}
  {\bibfnamefont {E.}~\bibnamefont {Thewalt}}, \bibinfo {author} {\bibfnamefont
  {A.}~\bibnamefont {Little}}, \bibinfo {author} {\bibfnamefont {J.~G.}\
  \bibnamefont {Analytis}}, \bibinfo {author} {\bibfnamefont {J.~E.}\
  \bibnamefont {Moore}},\ and\ \bibinfo {author} {\bibfnamefont
  {J.}~\bibnamefont {Orenstein}},\ }\bibfield  {title} {\bibinfo {title} {Giant
  anisotropic nonlinear optical response in transition metal monopnictide Weyl
  semimetals},\ }\href@noop {} {\bibfield  {journal} {\bibinfo  {journal}
  {Nat. Phys.}\ }\textbf {\bibinfo {volume} {13}},\ \bibinfo {pages} {350}
  (\bibinfo {year} {2017})}\BibitemShut {NoStop}%
\bibitem [{\citenamefont {Feng}\ \emph {et~al.}(2015)\citenamefont {Feng},
  \citenamefont {Guo}, \citenamefont {Zhou}, \citenamefont {Yao},\ and\
  \citenamefont {Niu}}]{feng2015large}%
  \BibitemOpen
  \bibfield  {author} {\bibinfo {author} {\bibfnamefont {W.}~\bibnamefont
  {Feng}}, \bibinfo {author} {\bibfnamefont {G.-Y.}\ \bibnamefont {Guo}},
  \bibinfo {author} {\bibfnamefont {J.}~\bibnamefont {Zhou}}, \bibinfo {author}
  {\bibfnamefont {Y.}~\bibnamefont {Yao}},\ and\ \bibinfo {author}
  {\bibfnamefont {Q.}~\bibnamefont {Niu}},\ }\bibfield  {title} {\bibinfo
  {title} {Large magneto-optical Kerr effect in noncollinear antiferromagnets
  Mn$_3$X (X= Rh, Ir, Pt)},\ }\href@noop {} {\bibfield  {journal} {\bibinfo
  {journal} {Phys. Rev. B}\ }\textbf {\bibinfo {volume} {92}},\ \bibinfo
  {pages} {144426} (\bibinfo {year} {2015})}\BibitemShut {NoStop}%
\bibitem [{\citenamefont {Higo}\ \emph {et~al.}(2018)\citenamefont {Higo},
  \citenamefont {Man}, \citenamefont {Gopman}, \citenamefont {Wu},
  \citenamefont {Koretsune}, \citenamefont {van’t Erve}, \citenamefont
  {Kabanov}, \citenamefont {Rees}, \citenamefont {Li}, \citenamefont {Suzuki}
  \emph {et~al.}}]{higo2018large}%
  \BibitemOpen
  \bibfield  {author} {\bibinfo {author} {\bibfnamefont {T.}~\bibnamefont
  {Higo}}, \bibinfo {author} {\bibfnamefont {H.}~\bibnamefont {Man}}, \bibinfo
  {author} {\bibfnamefont {D.~B.}\ \bibnamefont {Gopman}}, \bibinfo {author}
  {\bibfnamefont {L.}~\bibnamefont {Wu}}, \bibinfo {author} {\bibfnamefont
  {T.}~\bibnamefont {Koretsune}}, \bibinfo {author} {\bibfnamefont {O.~M.}\
  \bibnamefont {van’t Erve}}, \bibinfo {author} {\bibfnamefont {Y.~P.}\
  \bibnamefont {Kabanov}}, \bibinfo {author} {\bibfnamefont {D.}~\bibnamefont
  {Rees}}, \bibinfo {author} {\bibfnamefont {Y.}~\bibnamefont {Li}}, \bibinfo
  {author} {\bibfnamefont {M.-T.}\ \bibnamefont {Suzuki}},
  \bibinfo{author} {\bibfnamefont {S.}\ \bibnamefont {Patankar}}, \bibinfo
  {author} {\bibfnamefont {M.}\ \bibnamefont {Ikhlas}},
  \bibinfo{author} {\bibfnamefont {C.L.}\ \bibnamefont {Chien}},
  \bibinfo{author} {\bibfnamefont {R.}\ \bibnamefont {Arita}}, \bibinfo{author} {\bibfnamefont {R.D.}\ \bibnamefont {Shull}},
   \bibinfo{author} {\bibfnamefont {J.}\ \bibnamefont {Orenstein}},\ and\ \bibinfo{author} {\bibfnamefont {S.}\ \bibnamefont {Nakatsuji}},\ }\bibfield  {title} {\bibinfo {title} {Large magneto-optical Kerr effect and
  imaging of magnetic octupole domains in an antiferromagnetic metal},\
  }\href@noop {} {\bibfield  {journal} {\bibinfo  {journal} {Nat. Photonics}\
  }\textbf {\bibinfo {volume} {12}},\ \bibinfo {pages} {73} (\bibinfo {year}
  {2018})}\BibitemShut {NoStop}%
\bibitem [{\citenamefont {Xu}\ \emph {et~al.}(2015{\natexlab{a}})\citenamefont
  {Xu}, \citenamefont {Belopolski}, \citenamefont {Alidoust}, \citenamefont
  {Neupane}, \citenamefont {Bian}, \citenamefont {Zhang}, \citenamefont
  {Sankar}, \citenamefont {Chang}, \citenamefont {Yuan}, \citenamefont {Lee},
  \citenamefont {Huang}, \citenamefont {Zheng}, \citenamefont {Ma},
  \citenamefont {Sanchez}, \citenamefont {Wang}, \citenamefont {Bansil},
  \citenamefont {Chou}, \citenamefont {Shibayev}, \citenamefont {Lin},
  \citenamefont {Jia},\ and\ \citenamefont {Hasan}}]{Xu15}%
  \BibitemOpen
  \bibfield  {author} {\bibinfo {author} {\bibfnamefont {S.-Y.}\ \bibnamefont
  {Xu}}, \bibinfo {author} {\bibfnamefont {I.}~\bibnamefont {Belopolski}},
  \bibinfo {author} {\bibfnamefont {N.}~\bibnamefont {Alidoust}}, \bibinfo
  {author} {\bibfnamefont {M.}~\bibnamefont {Neupane}}, \bibinfo {author}
  {\bibfnamefont {G.}~\bibnamefont {Bian}}, \bibinfo {author} {\bibfnamefont
  {C.}~\bibnamefont {Zhang}}, \bibinfo {author} {\bibfnamefont
  {R.}~\bibnamefont {Sankar}}, \bibinfo {author} {\bibfnamefont
  {G.}~\bibnamefont {Chang}}, \bibinfo {author} {\bibfnamefont
  {Z.}~\bibnamefont {Yuan}}, \bibinfo {author} {\bibfnamefont {C.-C.}\
  \bibnamefont {Lee}}, \bibinfo {author} {\bibfnamefont {S.-M.}\ \bibnamefont
  {Huang}}, \bibinfo {author} {\bibfnamefont {H.}~\bibnamefont {Zheng}},
  \bibinfo {author} {\bibfnamefont {J.}~\bibnamefont {Ma}}, \bibinfo {author}
  {\bibfnamefont {D.~S.}\ \bibnamefont {Sanchez}}, \bibinfo {author}
  {\bibfnamefont {B.}~\bibnamefont {Wang}}, \bibinfo {author} {\bibfnamefont
  {A.}~\bibnamefont {Bansil}}, \bibinfo {author} {\bibfnamefont
  {F.}~\bibnamefont {Chou}}, \bibinfo {author} {\bibfnamefont {P.~P.}\
  \bibnamefont {Shibayev}}, \bibinfo {author} {\bibfnamefont {H.}~\bibnamefont
  {Lin}}, \bibinfo {author} {\bibfnamefont {S.}~\bibnamefont {Jia}},\ and\
  \bibinfo {author} {\bibfnamefont {M.}\ \bibnamefont {Zahid Hasan}},\ }\bibfield
  {title} {\bibinfo {title} {Discovery of a Weyl Fermion semimetal and
  topological Fermi arcs},\ }\href {https://doi.org/10.1126/science.aaa9297}
  {\bibfield  {journal} {\bibinfo  {journal} {Science}\ }\textbf {\bibinfo
  {volume} {349}},\ \bibinfo {pages} {613} (\bibinfo {year}
  {2015}{\natexlab{a}})}\ \BibitemShut
  {NoStop}%
\bibitem [{\citenamefont {Xu}\ \emph {et~al.}(2015{\natexlab{b}})\citenamefont
  {Xu}, \citenamefont {Alidoust}, \citenamefont {Belopolski}, \citenamefont
  {Yuan}, \citenamefont {Bian}, \citenamefont {Chang}, \citenamefont {Zheng},
  \citenamefont {Strocov}, \citenamefont {Sanchez}, \citenamefont {Chang},
  \citenamefont {Zhang}, \citenamefont {Mou}, \citenamefont {Wu}, \citenamefont
  {Huang}, \citenamefont {Lee}, \citenamefont {Huang}, \citenamefont {Wang},
  \citenamefont {Bansil}, \citenamefont {Jeng}, \citenamefont {Neupert},
  \citenamefont {Kaminski}, \citenamefont {Lin}, \citenamefont {Jia},\ and\
  \citenamefont {Zahid~Hasan}}]{Xu2015}%
  \BibitemOpen
  \bibfield  {author} {\bibinfo {author} {\bibfnamefont {S.-Y.}\ \bibnamefont
  {Xu}}, \bibinfo {author} {\bibfnamefont {N.}~\bibnamefont {Alidoust}},
  \bibinfo {author} {\bibfnamefont {I.}~\bibnamefont {Belopolski}}, \bibinfo
  {author} {\bibfnamefont {Z.}~\bibnamefont {Yuan}}, \bibinfo {author}
  {\bibfnamefont {G.}~\bibnamefont {Bian}}, \bibinfo {author} {\bibfnamefont
  {T.-R.}\ \bibnamefont {Chang}}, \bibinfo {author} {\bibfnamefont
  {H.}~\bibnamefont {Zheng}}, \bibinfo {author} {\bibfnamefont {V.~N.}\
  \bibnamefont {Strocov}}, \bibinfo {author} {\bibfnamefont {D.~S.}\
  \bibnamefont {Sanchez}}, \bibinfo {author} {\bibfnamefont {G.}~\bibnamefont
  {Chang}}, \bibinfo {author} {\bibfnamefont {C.}~\bibnamefont {Zhang}},
  \bibinfo {author} {\bibfnamefont {D.}~\bibnamefont {Mou}}, \bibinfo {author}
  {\bibfnamefont {Y.}~\bibnamefont {Wu}}, \bibinfo {author} {\bibfnamefont
  {L.}~\bibnamefont {Huang}}, \bibinfo {author} {\bibfnamefont {C.-C.}\
  \bibnamefont {Lee}}, \bibinfo {author} {\bibfnamefont {S.-M.}\ \bibnamefont
  {Huang}}, \bibinfo {author} {\bibfnamefont {B.}~\bibnamefont {Wang}},
  \bibinfo {author} {\bibfnamefont {A.}~\bibnamefont {Bansil}}, \bibinfo
  {author} {\bibfnamefont {H.-T.}\ \bibnamefont {Jeng}}, \bibinfo {author}
  {\bibfnamefont {T.}~\bibnamefont {Neupert}}, \bibinfo {author} {\bibfnamefont
  {A.}~\bibnamefont {Kaminski}}, \bibinfo {author} {\bibfnamefont
  {H.}~\bibnamefont {Lin}}, \bibinfo {author} {\bibfnamefont {S.}~\bibnamefont
  {Jia}},\ and\ \bibinfo {author} {\bibfnamefont {M.}~\bibnamefont
  {Zahid~Hasan}},\ }\bibfield  {title} {\bibinfo {title} {Discovery of a Weyl
  Fermion state with Fermi arcs in Niobium Arsenide},\ }\href
  {https://doi.org/10.1038/nphys3437} {\bibfield  {journal} {\bibinfo
  {journal} {Nat. Phys.}\ }\textbf {\bibinfo {volume} {11}},\ \bibinfo
  {pages} {748} (\bibinfo {year} {2015}{\natexlab{b}})}\BibitemShut {NoStop}%
\bibitem [{\citenamefont {Hasan}\ \emph {et~al.}(2017)\citenamefont {Hasan},
  \citenamefont {Xu}, \citenamefont {Belopolski},\ and\ \citenamefont
  {Huang}}]{hasan17}%
  \BibitemOpen
  \bibfield  {author} {\bibinfo {author} {\bibfnamefont {M.}\ \bibnamefont
  {Zahid Hasan}}, \bibinfo {author} {\bibfnamefont {S.-Y.}\ \bibnamefont {Xu}},
  \bibinfo {author} {\bibfnamefont {I.}~\bibnamefont {Belopolski}},\ and\
  \bibinfo {author} {\bibfnamefont {S.-M.}\ \bibnamefont {Huang}},\ }\bibfield
  {title} {\bibinfo {title} {Discovery of Weyl Fermion semimetals and
  topological Fermi arc states},\ }\href
  {https://doi.org/10.1146/annurev-conmatphys-031016-025225} {\bibfield
  {journal} {\bibinfo  {journal} {Annu. Rev. Condens. Matter Phys.}\
  }\textbf {\bibinfo {volume} {8}},\ \bibinfo {pages} {289} (\bibinfo {year}
  {2017})}\ \BibitemShut
  {NoStop}%
\bibitem [{\citenamefont {Huang}\ \emph {et~al.}(2015)\citenamefont {Huang},
  \citenamefont {Zhao}, \citenamefont {Long}, \citenamefont {Wang},
  \citenamefont {Chen}, \citenamefont {Yang}, \citenamefont {Liang},
  \citenamefont {Xue}, \citenamefont {Weng}, \citenamefont {Fang} \emph
  {et~al.}}]{huang2015observation}%
  \BibitemOpen
  \bibfield  {author} {\bibinfo {author} {\bibfnamefont {X.}~\bibnamefont
  {Huang}}, \bibinfo {author} {\bibfnamefont {L.}~\bibnamefont {Zhao}},
  \bibinfo {author} {\bibfnamefont {Y.}~\bibnamefont {Long}}, \bibinfo {author}
  {\bibfnamefont {P.}~\bibnamefont {Wang}}, \bibinfo {author} {\bibfnamefont
  {D.}~\bibnamefont {Chen}}, \bibinfo {author} {\bibfnamefont {Z.}~\bibnamefont
  {Yang}}, \bibinfo {author} {\bibfnamefont {H.}~\bibnamefont {Liang}},
  \bibinfo {author} {\bibfnamefont {M.}~\bibnamefont {Xue}}, \bibinfo {author}
  {\bibfnamefont {H.}~\bibnamefont {Weng}}, \bibinfo {author} {\bibfnamefont
  {Z.}~\bibnamefont {Fang}}, \bibinfo {author} {\bibfnamefont
  {X.}~\bibnamefont {Dai}},\ and\ \bibinfo {author} {\bibfnamefont
  {G.}~\bibnamefont {Chen}},\ }\bibfield  {title} {\bibinfo
  {title} {Observation of the chiral-anomaly-induced negative magnetoresistance
  in 3D Weyl semimetal TaAs},\ }\href@noop {} {\bibfield  {journal} {\bibinfo
  {journal} {Phys. Rev. X}\ }\textbf {\bibinfo {volume} {5}},\ \bibinfo
  {pages} {031023} (\bibinfo {year} {2015})}\BibitemShut {NoStop}%
\bibitem [{\citenamefont {Yan}\ and\ \citenamefont
  {Felser}(2017)}]{yan2017topological}%
  \BibitemOpen
  \bibfield  {author} {\bibinfo {author} {\bibfnamefont {B.}~\bibnamefont
  {Yan}}\ and\ \bibinfo {author} {\bibfnamefont {C.}~\bibnamefont {Felser}},\
  }\bibfield  {title} {\bibinfo {title} {Topological materials: Weyl
  semimetals},\ }\href@noop {} {\bibfield  {journal} {\bibinfo  {journal}
  {Annu. Rev. Condens. Matter Phys.}\ }\textbf {\bibinfo {volume}
  {8}},\ \bibinfo {pages} {337} (\bibinfo {year} {2017})}\BibitemShut {NoStop}%
\bibitem [{\citenamefont {Kar}\ and\ \citenamefont {Jayannavar}(2021)}]{AJR2P}%
  \BibitemOpen
  \bibfield  {author} {\bibinfo {author} {\bibfnamefont {S.}~\bibnamefont
  {Kar}}\ and\ \bibinfo {author} {\bibfnamefont {A.}~\bibnamefont
  {Jayannavar}},\ }\bibfield  {title} {\bibinfo {title} {A primer on Weyl
  semimetals: Down the discovery of topological phases},\ }\href@noop {}
  {\bibfield  {journal} {\bibinfo  {journal} {Asian J. Res. Rev. Phys.}\ }\textbf {\bibinfo {volume} {4}},\ \bibinfo
  {pages} {34-35} (\bibinfo {year} {2021})}\BibitemShut {NoStop}%
\bibitem [{\citenamefont {Belopolski}\ \emph {et~al.}(2019)\citenamefont
  {Belopolski}, \citenamefont {Manna}, \citenamefont {Sanchez}, \citenamefont
  {Chang}, \citenamefont {Ernst}, \citenamefont {Yin}, \citenamefont {Zhang},
  \citenamefont {Cochran}, \citenamefont {Shumiya}, \citenamefont {Zheng} \emph
  {et~al.}}]{belopolski2019discovery}%
  \BibitemOpen
  \bibfield  {author} {\bibinfo {author} {\bibfnamefont {I.}~\bibnamefont
  {Belopolski}}, \bibinfo {author} {\bibfnamefont {K.}~\bibnamefont {Manna}},
  \bibinfo {author} {\bibfnamefont {D.~S.}\ \bibnamefont {Sanchez}}, \bibinfo
  {author} {\bibfnamefont {G.}~\bibnamefont {Chang}}, \bibinfo {author}
  {\bibfnamefont {B.}~\bibnamefont {Ernst}}, \bibinfo {author} {\bibfnamefont
  {J.}~\bibnamefont {Yin}}, \bibinfo {author} {\bibfnamefont {S.~S.}\
  \bibnamefont {Zhang}}, \bibinfo {author} {\bibfnamefont {T.}~\bibnamefont
  {Cochran}}, \bibinfo {author} {\bibfnamefont {N.}~\bibnamefont {Shumiya}},
  \bibinfo {author} {\bibfnamefont {H.}~\bibnamefont {Zheng}}, \bibinfo {author} {\bibfnamefont {B.}~\bibnamefont {Singh}},\bibinfo {author} {\bibfnamefont {G.}~\bibnamefont {Bian}},
  \bibinfo {author} {\bibfnamefont {D.}~\bibnamefont {Multer}},
  \bibinfo {author} {\bibfnamefont {M.}~\bibnamefont {Litskevich}},
  \bibinfo {author} {\bibfnamefont {X.}~\bibnamefont {Zhou}},
  \bibinfo {author} {\bibfnamefont {S.~M.}~\bibnamefont {Huang}},
  \bibinfo {author} {\bibfnamefont {B.}~\bibnamefont {Wang}},
  \bibinfo {author} {\bibfnamefont {T.R.}~\bibnamefont {Chang}},
  \bibinfo {author} {\bibfnamefont {S.-Y.}~\bibnamefont {Xu}},
  \bibinfo {author} {\bibfnamefont {A.}~\bibnamefont {Bansal}},
  \bibinfo {author} {\bibfnamefont {C.}~\bibnamefont {Felser}},
  \bibinfo {author} {\bibfnamefont {H.}~\bibnamefont {Lin}},\ and\ \bibinfo {author} {\bibfnamefont {M.}~\bibnamefont {Zahid Hasan}},\ }\bibfield  {title} {\bibinfo
  {title} {Discovery of topological Weyl Fermion lines and drumhead surface states in a room temperature magnet},\ }\href@noop {} {\bibfield  {journal} {\bibinfo  {journal} {Science}\ }\textbf {\bibinfo
  {volume} {365}},\ \bibinfo {pages} {1278} (\bibinfo {year}
  {2019})}\BibitemShut {NoStop}%
\bibitem [{\citenamefont {Liu}\ \emph {et~al.}(2017)\citenamefont {Liu},
  \citenamefont {Li}, \citenamefont {Cui}, \citenamefont {Deng},\ and\
  \citenamefont {Tao}}]{liu2017nonmagnetic}%
  \BibitemOpen
  \bibfield  {author} {\bibinfo {author} {\bibfnamefont {X.}~\bibnamefont
  {Liu}}, \bibinfo {author} {\bibfnamefont {L.}~\bibnamefont {Li}}, \bibinfo
  {author} {\bibfnamefont {Y.}~\bibnamefont {Cui}}, \bibinfo {author}
  {\bibfnamefont {J.}~\bibnamefont {Deng}},\ and\ \bibinfo {author}
  {\bibfnamefont {X.}~\bibnamefont {Tao}},\ }\bibfield  {title} {\bibinfo
  {title} {A nonmagnetic topological Weyl semimetal in quaternary Heusler
  compound CrAlTiV},\ }\href@noop {} {\bibfield  {journal} {\bibinfo  {journal}
  {Appl. Phys. Lett.}\ }\textbf {\bibinfo {volume} {111}},\ \bibinfo
  {pages} {122104} (\bibinfo {year} {2017})}\BibitemShut {NoStop}%
\bibitem [{\citenamefont {Manna}\ \emph {et~al.}(2018)\citenamefont {Manna},
  \citenamefont {Muechler}, \citenamefont {Kao}, \citenamefont {Stinshoff},
  \citenamefont {Zhang}, \citenamefont {Gooth}, \citenamefont {Kumar},
  \citenamefont {Kreiner}, \citenamefont {Koepernik}, \citenamefont {Car} \emph
  {et~al.}}]{manna2018colossal}%
  \BibitemOpen
  \bibfield  {author} {\bibinfo {author} {\bibfnamefont {K.}~\bibnamefont
  {Manna}}, \bibinfo {author} {\bibfnamefont {L.}~\bibnamefont {Muechler}},
  \bibinfo {author} {\bibfnamefont {T.-H.}\ \bibnamefont {Kao}}, \bibinfo
  {author} {\bibfnamefont {R.}~\bibnamefont {Stinshoff}}, \bibinfo {author}
  {\bibfnamefont {Y.}~\bibnamefont {Zhang}}, \bibinfo {author} {\bibfnamefont
  {J.}~\bibnamefont {Gooth}}, \bibinfo {author} {\bibfnamefont
  {N.}~\bibnamefont {Kumar}}, \bibinfo {author} {\bibfnamefont
  {G.}~\bibnamefont {Kreiner}}, \bibinfo {author} {\bibfnamefont
  {K.}~\bibnamefont {Koepernik}}, \bibinfo {author} {\bibfnamefont {R.}~\bibnamefont{Car}}, \bibinfo {author} {\bibfnamefont{J.}~\bibnamefont{K{\"u}bler}}, \bibinfo {author} {\bibfnamefont{G.H.}~\bibnamefont {Fecher}}, \bibinfo {author} {\bibfnamefont{C.}~\bibnamefont {Shekhar}}, \bibinfo {author} {\bibfnamefont{Y.}~\bibnamefont {Sun}},\ and\ \bibinfo {author} {\bibfnamefont{C.}~\bibnamefont {Felser}},\ }\bibfield  {title} {\bibinfo {title} {From colossal to zero: controlling the anomalous Hall effect in
  magnetic Heusler compounds via Berry curvature design},\ }\href@noop {}
  {\bibfield  {journal} {\bibinfo  {journal} {Phys. Rev. X}\ }\textbf
  {\bibinfo {volume} {8}},\ \bibinfo {pages} {041045} (\bibinfo {year}
  {2018})}\BibitemShut {NoStop}%
\bibitem [{\citenamefont {Ernst}\ \emph {et~al.}(2019)\citenamefont {Ernst},
  \citenamefont {Sahoo}, \citenamefont {Sun}, \citenamefont {Nayak},
  \citenamefont {M\"uchler}, \citenamefont {Nayak}, \citenamefont {Kumar},
  \citenamefont {Gayles}, \citenamefont {Markou}, \citenamefont {Fecher},\ and\
  \citenamefont {Felser}}]{ernst2019anomalous}%
  \BibitemOpen
  \bibfield  {author} {\bibinfo {author} {\bibfnamefont {B.}~\bibnamefont
  {Ernst}}, \bibinfo {author} {\bibfnamefont {R.}~\bibnamefont {Sahoo}},
  \bibinfo {author} {\bibfnamefont {Y.}~\bibnamefont {Sun}}, \bibinfo {author}
  {\bibfnamefont {J.}~\bibnamefont {Nayak}}, \bibinfo {author} {\bibfnamefont
  {L.}~\bibnamefont {M\"uchler}}, \bibinfo {author} {\bibfnamefont {A.~K.}\
  \bibnamefont {Nayak}}, \bibinfo {author} {\bibfnamefont {N.}~\bibnamefont
  {Kumar}}, \bibinfo {author} {\bibfnamefont {J.}~\bibnamefont {Gayles}},
  \bibinfo {author} {\bibfnamefont {A.}~\bibnamefont {Markou}}, \bibinfo
  {author} {\bibfnamefont {G.~H.}\ \bibnamefont {Fecher}},\ and\ \bibinfo
  {author} {\bibfnamefont {C.}~\bibnamefont {Felser}},\ }\bibfield  {title}
  {\bibinfo {title} {Anomalous Hall effect and the role of Berry curvature in Co$_2$TiSn Heusler films},\ }\href
  {https://doi.org/10.1103/PhysRevB.100.054445} {\bibfield  {journal} {\bibinfo
   {journal} {Phys. Rev. B}\ }\textbf {\bibinfo {volume} {100}},\ \bibinfo
  {pages} {054445} (\bibinfo {year} {2019})}\BibitemShut {NoStop}%
\bibitem [{\citenamefont {Dulal}\ \emph {et~al.}(2019)\citenamefont {Dulal},
  \citenamefont {Dahal}, \citenamefont {Forbes}, \citenamefont {Bhattarai},
  \citenamefont {Pegg},\ and\ \citenamefont {Philip}}]{dulal2019weak}%
  \BibitemOpen
  \bibfield  {author} {\bibinfo {author} {\bibfnamefont {R.~P.}\ \bibnamefont
  {Dulal}}, \bibinfo {author} {\bibfnamefont {B.~R.}\ \bibnamefont {Dahal}},
  \bibinfo {author} {\bibfnamefont {A.}~\bibnamefont {Forbes}}, \bibinfo
  {author} {\bibfnamefont {N.}~\bibnamefont {Bhattarai}}, \bibinfo {author}
  {\bibfnamefont {I.~L.}\ \bibnamefont {Pegg}},\ and\ \bibinfo {author}
  {\bibfnamefont {J.}~\bibnamefont {Philip}},\ }\bibfield  {title} {\bibinfo
  {title} {Weak localization and small anomalous Hall conductivity in
  ferromagnetic Weyl semimetal Co$_2$TiGe},\ }\href@noop {} {\bibfield
  {journal} {\bibinfo  {journal} {Sci. Rep.}\ }\textbf {\bibinfo
  {volume} {9}},\ \bibinfo {pages} {3342} (\bibinfo {year} {2019})}\BibitemShut
  {NoStop}%
\bibitem [{\citenamefont {Nakajima}\ \emph {et~al.}(2015)\citenamefont
  {Nakajima}, \citenamefont {Hu}, \citenamefont {Kirshenbaum}, \citenamefont
  {Hughes}, \citenamefont {Syers}, \citenamefont {Wang}, \citenamefont {Wang},
  \citenamefont {Wang}, \citenamefont {Saha}, \citenamefont {Pratt} \emph
  {et~al.}}]{nakajima2015topological}%
  \BibitemOpen
  \bibfield  {author} {\bibinfo {author} {\bibfnamefont {Y.}~\bibnamefont
  {Nakajima}}, \bibinfo {author} {\bibfnamefont {R.}~\bibnamefont {Hu}},
  \bibinfo {author} {\bibfnamefont {K.}~\bibnamefont {Kirshenbaum}}, \bibinfo
  {author} {\bibfnamefont {A.}~\bibnamefont {Hughes}}, \bibinfo {author}
  {\bibfnamefont {P.}~\bibnamefont {Syers}}, \bibinfo {author} {\bibfnamefont
  {X.}~\bibnamefont {Wang}}, \bibinfo {author} {\bibfnamefont {K.}~\bibnamefont
  {Wang}}, \bibinfo {author} {\bibfnamefont {R.}~\bibnamefont {Wang}}, \bibinfo
  {author} {\bibfnamefont {S.~R.}\ \bibnamefont {Saha}}, \bibinfo {author}
  {\bibfnamefont {D.}~\bibnamefont {Pratt}}, \bibinfo {author}
  {\bibfnamefont {J.W.}~\bibnamefont {Lynn}},\ and\ \bibinfo {author}
  {\bibfnamefont {J.}~\bibnamefont {Paglione}},\  }\bibfield{title} {\bibinfo {title} {Topological RPdBi half-Heusler semimetals: A new
  family of noncentrosymmetric magnetic superconductors},\ }\href@noop {}
  {\bibfield  {journal} {\bibinfo  {journal} {Sci. Adv.}\ }\textbf
  {\bibinfo {volume} {1}},\ \bibinfo {pages} {e1500242} (\bibinfo {year}
  {2015})}\BibitemShut {NoStop}%
\bibitem [{\citenamefont {Kim}\ \emph {et~al.}(2018)\citenamefont {Kim},
  \citenamefont {Wang}, \citenamefont {Nakajima}, \citenamefont {Hu},
  \citenamefont {Ziemak}, \citenamefont {Syers}, \citenamefont {Wang},
  \citenamefont {Hodovanets}, \citenamefont {Denlinger}, \citenamefont {Brydon}
  \emph {et~al.}}]{kim2018beyond}%
  \BibitemOpen
  \bibfield  {author} {\bibinfo {author} {\bibfnamefont {H.}~\bibnamefont
  {Kim}}, \bibinfo {author} {\bibfnamefont {K.}~\bibnamefont {Wang}}, \bibinfo
  {author} {\bibfnamefont {Y.}~\bibnamefont {Nakajima}}, \bibinfo {author}
  {\bibfnamefont {R.}~\bibnamefont {Hu}}, \bibinfo {author} {\bibfnamefont
  {S.}~\bibnamefont {Ziemak}}, \bibinfo {author} {\bibfnamefont
  {P.}~\bibnamefont {Syers}}, \bibinfo {author} {\bibfnamefont
  {L.}~\bibnamefont {Wang}}, \bibinfo {author} {\bibfnamefont {H.}~\bibnamefont
  {Hodovanets}}, \bibinfo {author} {\bibfnamefont {J.~D.}\ \bibnamefont
  {Denlinger}}, \bibinfo {author} {\bibfnamefont {P.~M.}\ \bibnamefont
  {Brydon}}, \bibinfo {author} {\bibfnamefont {D.~F.}\ \bibnamefont
  {Agterberg}}, \bibinfo {author} {\bibfnamefont {M.~A.}\ \bibnamefont
  {Tanatar}}, \bibinfo {author} {\bibfnamefont {R.}\ \bibnamefont
  {Prozorov}},\ and\ \bibinfo {author} {\bibfnamefont {J.}\ \bibnamefont
  {Paglione}},\ }\bibfield  {title} {\bibinfo {title} {Beyond triplet: Unconventional superconductivity in a spin-3/2 topological
  semimetal},\ }\href@noop {} {\bibfield  {journal} {\bibinfo  {journal}
  {Sci. Adv.}\ }\textbf {\bibinfo {volume} {4}},\ \bibinfo {pages}
  {eaao4513} (\bibinfo {year} {2018})}\BibitemShut {NoStop}%
\bibitem [{\citenamefont {Chang}\ \emph {et~al.}(2017)\citenamefont {Chang},
  \citenamefont {Xu}, \citenamefont {Zhou}, \citenamefont {Huang},
  \citenamefont {Singh}, \citenamefont {Wang}, \citenamefont {Belopolski},
  \citenamefont {Yin}, \citenamefont {Zhang}, \citenamefont {Bansil},
  \citenamefont {Lin},\ and\ \citenamefont {Hasan}}]{chang2017topological}%
  \BibitemOpen
  \bibfield  {author} {\bibinfo {author} {\bibfnamefont {G.}~\bibnamefont
  {Chang}}, \bibinfo {author} {\bibfnamefont {S.-Y.}\ \bibnamefont {Xu}},
  \bibinfo {author} {\bibfnamefont {X.}~\bibnamefont {Zhou}}, \bibinfo {author}
  {\bibfnamefont {S.-M.}\ \bibnamefont {Huang}}, \bibinfo {author}
  {\bibfnamefont {B.}~\bibnamefont {Singh}}, \bibinfo {author} {\bibfnamefont
  {B.}~\bibnamefont {Wang}}, \bibinfo {author} {\bibfnamefont {I.}~\bibnamefont
  {Belopolski}}, \bibinfo {author} {\bibfnamefont {J.}~\bibnamefont {Yin}},
  \bibinfo {author} {\bibfnamefont {S.}~\bibnamefont {Zhang}}, \bibinfo
  {author} {\bibfnamefont {A.}~\bibnamefont {Bansil}}, \bibinfo {author}
  {\bibfnamefont {H.}~\bibnamefont {Lin}},\ and\ \bibinfo {author}
  {\bibfnamefont {M.}\ \bibnamefont {Zahid Hasan}},\ }\bibfield  {title} {\bibinfo
  {title} {Topological hopf and chain link semimetal states and their
  application to Co$_2$MnGa},\ }\href@noop
  {} {\bibfield  {journal} {\bibinfo  {journal} {Phys. Rev. Lett.}\ }\textbf
  {\bibinfo {volume} {119}},\ \bibinfo {pages} {156401} (\bibinfo {year}
  {2017})}\BibitemShut {NoStop}%
\bibitem [{\citenamefont {Reichlova}\ \emph {et~al.}(2018)\citenamefont
  {Reichlova}, \citenamefont {Schlitz}, \citenamefont {Beckert}, \citenamefont
  {Swekis}, \citenamefont {Markou}, \citenamefont {Chen}, \citenamefont
  {Kriegner}, \citenamefont {Fabretti}, \citenamefont {Hyeon~Park},
  \citenamefont {Niemann} \emph {et~al.}}]{reichlova2018large}%
  \BibitemOpen
  \bibfield  {author} {\bibinfo {author} {\bibfnamefont {H.}~\bibnamefont
  {Reichlova}}, \bibinfo {author} {\bibfnamefont {R.}~\bibnamefont {Schlitz}},
  \bibinfo {author} {\bibfnamefont {S.}~\bibnamefont {Beckert}}, \bibinfo
  {author} {\bibfnamefont {P.}~\bibnamefont {Swekis}}, \bibinfo {author}
  {\bibfnamefont {A.}~\bibnamefont {Markou}}, \bibinfo {author} {\bibfnamefont
  {Y.-C.}\ \bibnamefont {Chen}}, \bibinfo {author} {\bibfnamefont
  {D.}~\bibnamefont {Kriegner}}, \bibinfo {author} {\bibfnamefont
  {S.}~\bibnamefont {Fabretti}}, \bibinfo {author} {\bibfnamefont
  {G.}~\bibnamefont {Hyeon~Park}}, \bibinfo {author} {\bibfnamefont
  {A.}~\bibnamefont {Niemann}}, \bibinfo {author} {\bibfnamefont
  {S.}~\bibnamefont {Sudheendra}}, \bibinfo {author} {\bibfnamefont
  {A.}~\bibnamefont {Thomas}}, \bibinfo {author} {\bibfnamefont
  {K.}~\bibnamefont {Nielsch}}, \bibinfo {author} {\bibfnamefont
  {C.}~\bibnamefont {Felser}},\ and\ \bibinfo {author} {\bibfnamefont
  {S. T. B.}~\bibnamefont {Goennenwein}},\ }\bibfield  {title} {\bibinfo
  {title} {Large anomalous Nernst effect in thin films of the Weyl semimetal
  Co$_2$MnGa},\ }\href@noop {} {\bibfield  {journal} {\bibinfo  {journal} {Appl. Phys. Lett.}\ }\textbf {\bibinfo {volume} {113}},\ \bibinfo {pages}
  {212405} (\bibinfo {year} {2018})}\BibitemShut {NoStop}%
\bibitem [{\citenamefont {Guin}\ \emph {et~al.}(2019)\citenamefont {Guin},
  \citenamefont {Manna}, \citenamefont {Noky}, \citenamefont {Watzman},
  \citenamefont {Fu}, \citenamefont {Kumar}, \citenamefont {Schnelle},
  \citenamefont {Shekhar}, \citenamefont {Sun}, \citenamefont {Gooth} \emph
  {et~al.}}]{guin2019anomalous}%
  \BibitemOpen
  \bibfield  {author} {\bibinfo {author} {\bibfnamefont {S.~N.}\ \bibnamefont
  {Guin}}, \bibinfo {author} {\bibfnamefont {K.}~\bibnamefont {Manna}},
  \bibinfo {author} {\bibfnamefont {J.}~\bibnamefont {Noky}}, \bibinfo {author}
  {\bibfnamefont {S.~J.}\ \bibnamefont {Watzman}}, \bibinfo {author}
  {\bibfnamefont {C.}~\bibnamefont {Fu}}, \bibinfo {author} {\bibfnamefont
  {N.}~\bibnamefont {Kumar}}, \bibinfo {author} {\bibfnamefont
  {W.}~\bibnamefont {Schnelle}}, \bibinfo {author} {\bibfnamefont
  {C.}~\bibnamefont {Shekhar}}, \bibinfo {author} {\bibfnamefont
  {Y.}~\bibnamefont {Sun}}, \bibinfo {author} {\bibfnamefont {J.}~\bibnamefont
  {Gooth}},\ and\  \bibinfo {author} {\bibfnamefont {C.}~\bibnamefont
  {Felser}},\ }\bibfield  {title} {\bibinfo {title} {Anomalous
  Nernst effect beyond the magnetization scaling relation in the ferromagnetic
  Heusler compound Co$_2$MnGa},\ }\href@noop {} {\bibfield  {journal} {\bibinfo
  {journal} {NPG Asia Mater.}\ }\textbf {\bibinfo {volume} {11}},\ \bibinfo
  {pages} {16} (\bibinfo {year} {2019})}\BibitemShut {NoStop}%
\bibitem [{\citenamefont {Sato}\ \emph {et~al.}(2019)\citenamefont {Sato},
  \citenamefont {Kokado}, \citenamefont {Tsujikawa}, \citenamefont {Ogawa},
  \citenamefont {Kosaka}, \citenamefont {Shirai},\ and\ \citenamefont
  {Tsunoda}}]{takashi2019signs}%
  \BibitemOpen
  \bibfield  {author} {\bibinfo {author} {\bibfnamefont {T.}~\bibnamefont
  {Sato}}, \bibinfo {author} {\bibfnamefont {S.}~\bibnamefont {Kokado}},
  \bibinfo {author} {\bibfnamefont {M.}~\bibnamefont {Tsujikawa}}, \bibinfo
  {author} {\bibfnamefont {T.}~\bibnamefont {Ogawa}}, \bibinfo {author}
  {\bibfnamefont {S.}~\bibnamefont {Kosaka}}, \bibinfo {author} {\bibfnamefont
  {M.}~\bibnamefont {Shirai}},\ and\ \bibinfo {author} {\bibfnamefont
  {M.}~\bibnamefont {Tsunoda}},\ }\bibfield  {title} {\bibinfo {title} {Signs
  of anisotropic magnetoresistance in Co$_2$MnGa Heusler alloy epitaxial thin
  films based on current direction},\ }\href@noop {} {\bibfield  {journal}
  {\bibinfo  {journal} {Appl. Phys. Express}\ }\textbf {\bibinfo {volume}
  {12}},\ \bibinfo {pages} {103005} (\bibinfo {year} {2019})}\BibitemShut
  {NoStop}%
\bibitem [{\citenamefont {Markou}\ \emph {et~al.}(2019)\citenamefont {Markou},
  \citenamefont {Kriegner}, \citenamefont {Gayles}, \citenamefont {Zhang},
  \citenamefont {Chen}, \citenamefont {Ernst}, \citenamefont {Lai},
  \citenamefont {Schnelle}, \citenamefont {Chu}, \citenamefont {Sun} \emph
  {et~al.}}]{markou2019thickness}%
  \BibitemOpen
  \bibfield  {author} {\bibinfo {author} {\bibfnamefont {A.}~\bibnamefont
  {Markou}}, \bibinfo {author} {\bibfnamefont {D.}~\bibnamefont {Kriegner}},
  \bibinfo {author} {\bibfnamefont {J.}~\bibnamefont {Gayles}}, \bibinfo
  {author} {\bibfnamefont {L.}~\bibnamefont {Zhang}}, \bibinfo {author}
  {\bibfnamefont {Y.-C.}\ \bibnamefont {Chen}}, \bibinfo {author}
  {\bibfnamefont {B.}~\bibnamefont {Ernst}}, \bibinfo {author} {\bibfnamefont
  {Y.-H.}\ \bibnamefont {Lai}}, \bibinfo {author} {\bibfnamefont
  {W.}~\bibnamefont {Schnelle}}, \bibinfo {author} {\bibfnamefont {Y.-H.}\
  \bibnamefont {Chu}}, \bibinfo {author} {\bibfnamefont {Y.}~\bibnamefont
  {Sun}},\ and\ \bibinfo {author} {\bibfnamefont {C.}~\bibnamefont
  {Felser}},\ }\bibfield  {title} {\bibinfo {title} {Thickness
  dependence of the anomalous Hall effect in thin films of the topological
  semimetal Co$_2$MnGa},\ }\href@noop {} {\bibfield  {journal} {\bibinfo
  {journal} {Phys. Rev. B}\ }\textbf {\bibinfo {volume} {100}},\ \bibinfo
  {pages} {054422} (\bibinfo {year} {2019})}\BibitemShut {NoStop}%
\bibitem [{\citenamefont {Park}\ \emph {et~al.}(2020)\citenamefont {Park},
  \citenamefont {Reichlova}, \citenamefont {Schlitz}, \citenamefont {Lammel},
  \citenamefont {Markou}, \citenamefont {Swekis}, \citenamefont {Ritzinger},
  \citenamefont {Kriegner}, \citenamefont {Noky}, \citenamefont {Gayles},
  \citenamefont {Sun}, \citenamefont {Felser}, \citenamefont {Nielsch},
  \citenamefont {Goennenwein},\ and\ \citenamefont
  {Thomas}}]{park2020thickness}%
  \BibitemOpen
  \bibfield  {author} {\bibinfo {author} {\bibfnamefont {G.-H.}\ \bibnamefont
  {Park}}, \bibinfo {author} {\bibfnamefont {H.}~\bibnamefont {Reichlova}},
  \bibinfo {author} {\bibfnamefont {R.}~\bibnamefont {Schlitz}}, \bibinfo
  {author} {\bibfnamefont {M.}~\bibnamefont {Lammel}}, \bibinfo {author}
  {\bibfnamefont {A.}~\bibnamefont {Markou}}, \bibinfo {author} {\bibfnamefont
  {P.}~\bibnamefont {Swekis}}, \bibinfo {author} {\bibfnamefont
  {P.}~\bibnamefont {Ritzinger}}, \bibinfo {author} {\bibfnamefont
  {D.}~\bibnamefont {Kriegner}}, \bibinfo {author} {\bibfnamefont
  {J.}~\bibnamefont {Noky}}, \bibinfo {author} {\bibfnamefont {J.}~\bibnamefont
  {Gayles}}, \bibinfo {author} {\bibfnamefont {Y.}~\bibnamefont {Sun}},
  \bibinfo {author} {\bibfnamefont {C.}~\bibnamefont {Felser}}, \bibinfo
  {author} {\bibfnamefont {K.}~\bibnamefont {Nielsch}}, \bibinfo {author}
  {\bibfnamefont {S.~T.~B.}\ \bibnamefont {Goennenwein}},\ and\ \bibinfo
  {author} {\bibfnamefont {A.}~\bibnamefont {Thomas}},\ }\bibfield  {title}
  {\bibinfo {title} {Thickness dependence of the anomalous nernst effect and
  the mott relation of Weyl semimetal Co$_2$MnGa thin
  films},\ }\href {https://doi.org/10.1103/PhysRevB.101.060406} {\bibfield
  {journal} {\bibinfo  {journal} {Phys. Rev. B}\ }\textbf {\bibinfo {volume}
  {101}},\ \bibinfo {pages} {060406} (\bibinfo {year} {2020})}\BibitemShut
  {NoStop}%
\bibitem [{\citenamefont {K{\"u}bler}\ and\ \citenamefont
  {Felser}(2016)}]{kubler2016weyl}%
  \BibitemOpen
  \bibfield  {author} {\bibinfo {author} {\bibfnamefont {J.}~\bibnamefont
  {K{\"u}bler}}\ and\ \bibinfo {author} {\bibfnamefont {C.}~\bibnamefont
  {Felser}},\ }\bibfield  {title} {\bibinfo {title} {Weyl points in the
  ferromagnetic Heusler compound Co$_2$MnAl},\ }\href@noop {} {\bibfield
  {journal} {\bibinfo  {journal} {Europhys. Lett.}\ }\textbf
  {\bibinfo {volume} {114}},\ \bibinfo {pages} {47005} (\bibinfo {year}
  {2016})}\BibitemShut {NoStop}%
\bibitem [{\citenamefont {Chang}\ \emph {et~al.}(2016)\citenamefont {Chang},
  \citenamefont {Xu}, \citenamefont {Zheng}, \citenamefont {Singh},
  \citenamefont {Hsu}, \citenamefont {Bian}, \citenamefont {Alidoust},
  \citenamefont {Belopolski}, \citenamefont {Sanchez}, \citenamefont {Zhang}
  \emph {et~al.}}]{chang2016room}%
  \BibitemOpen
  \bibfield  {author} {\bibinfo {author} {\bibfnamefont {G.}~\bibnamefont
  {Chang}}, \bibinfo {author} {\bibfnamefont {S.-Y.}\ \bibnamefont {Xu}},
  \bibinfo {author} {\bibfnamefont {H.}~\bibnamefont {Zheng}}, \bibinfo
  {author} {\bibfnamefont {B.}~\bibnamefont {Singh}}, \bibinfo {author}
  {\bibfnamefont {C.-H.}\ \bibnamefont {Hsu}}, \bibinfo {author} {\bibfnamefont
  {G.}~\bibnamefont {Bian}}, \bibinfo {author} {\bibfnamefont {N.}~\bibnamefont
  {Alidoust}}, \bibinfo {author} {\bibfnamefont {I.}~\bibnamefont
  {Belopolski}}, \bibinfo {author} {\bibfnamefont {D.~S.}\ \bibnamefont
  {Sanchez}}, \bibinfo {author} {\bibfnamefont {S.}~\bibnamefont {Zhang}},
  \bibinfo {author} {\bibfnamefont {H.}~\bibnamefont {Lin}},\ and\ \bibinfo {author} {\bibfnamefont {M.}~\bibnamefont {Zahid Hasan}}, \ }\bibfield {title} {\bibinfo {title} {Room-temperature
  magnetic topological Weyl Fermion and nodal line semimetal states in
  half-metallic Heusler Co$_2$TiX (X = Si, Ge, or Sn)},\ }\href@noop {}
  {\bibfield  {journal} {\bibinfo  {journal} {Sci. Rep.}\ }\textbf
  {\bibinfo {volume} {6}},\ \bibinfo {pages} {38839} (\bibinfo {year}
  {2016})}\BibitemShut {NoStop}%
\bibitem [{\citenamefont {Wang}\ \emph {et~al.}(2016)\citenamefont {Wang},
  \citenamefont {Vergniory}, \citenamefont {Kushwaha}, \citenamefont
  {Hirschberger}, \citenamefont {Chulkov}, \citenamefont {Ernst}, \citenamefont
  {Ong}, \citenamefont {Cava},\ and\ \citenamefont {Bernevig}}]{wang2016time}%
  \BibitemOpen
  \bibfield  {author} {\bibinfo {author} {\bibfnamefont {Z.}~\bibnamefont
  {Wang}}, \bibinfo {author} {\bibfnamefont {M. G.}~\bibnamefont {Vergniory}},
  \bibinfo {author} {\bibfnamefont {S.}~\bibnamefont {Kushwaha}}, \bibinfo
  {author} {\bibfnamefont {M.}~\bibnamefont {Hirschberger}}, \bibinfo {author}
  {\bibfnamefont {E. V.}~\bibnamefont {Chulkov}}, \bibinfo {author} {\bibfnamefont
  {A.}~\bibnamefont {Ernst}}, \bibinfo {author} {\bibfnamefont {N.~P.}\
  \bibnamefont {Ong}}, \bibinfo {author} {\bibfnamefont {R.~J.}\ \bibnamefont
  {Cava}},\ and\ \bibinfo {author} {\bibfnamefont {B.~A.}\ \bibnamefont
  {Bernevig}},\ }\bibfield  {title} {\bibinfo {title} {Time-reversal-breaking
  Weyl Fermions in magnetic Heusler alloys},\ }\href@noop {} {\bibfield
  {journal} {\bibinfo  {journal} {Phys. Rev. Lett.}\ }\textbf {\bibinfo
  {volume} {117}},\ \bibinfo {pages} {236401} (\bibinfo {year}
  {2016})}\BibitemShut {NoStop}%
\bibitem [{\citenamefont {Chadov}\ \emph {et~al.}(2017)\citenamefont {Chadov},
  \citenamefont {Wu}, \citenamefont {Felser},\ and\ \citenamefont
  {Galanakis}}]{chadov2017stability}%
  \BibitemOpen
  \bibfield  {author} {\bibinfo {author} {\bibfnamefont {S.}~\bibnamefont
  {Chadov}}, \bibinfo {author} {\bibfnamefont {S.-C.}\ \bibnamefont {Wu}},
  \bibinfo {author} {\bibfnamefont {C.}~\bibnamefont {Felser}},\ and\ \bibinfo
  {author} {\bibfnamefont {I.}~\bibnamefont {Galanakis}},\ }\bibfield  {title}
  {\bibinfo {title} {Stability of Weyl points in magnetic half-metallic Heusler
  compounds},\ }\href@noop {} {\bibfield  {journal} {\bibinfo  {journal}
  {Phys. Rev. B}\ }\textbf {\bibinfo {volume} {96}},\ \bibinfo {pages}
  {024435} (\bibinfo {year} {2017})}\BibitemShut {NoStop}%
\bibitem [{\citenamefont {Kushwaha}\ \emph {et~al.}(2018)\citenamefont
  {Kushwaha}, \citenamefont {Wang}, \citenamefont {Kong},\ and\ \citenamefont
  {Cava}}]{kushwaha2018magnetic}%
  \BibitemOpen
  \bibfield  {author} {\bibinfo {author} {\bibfnamefont {S.~K.}\ \bibnamefont
  {Kushwaha}}, \bibinfo {author} {\bibfnamefont {Z.}~\bibnamefont {Wang}},
  \bibinfo {author} {\bibfnamefont {T.}~\bibnamefont {Kong}},\ and\ \bibinfo
  {author} {\bibfnamefont {R.~J.}\ \bibnamefont {Cava}},\ }\bibfield  {title}
  {\bibinfo {title} {Magnetic and electronic properties of the Cu-substituted
  Weyl semimetal candidate ZrCo$_2$Sn},\ }\href@noop {} {\bibfield  {journal}
  {\bibinfo  {journal} {J. Phys.: Condens. Matter}\ }\textbf
  {\bibinfo {volume} {30}},\ \bibinfo {pages} {075701} (\bibinfo {year}
  {2018})}\BibitemShut {NoStop}%
\bibitem [{\citenamefont {Yang}\ \emph {et~al.}(2019)\citenamefont {Yang},
  \citenamefont {Gu}, \citenamefont {Yi}, \citenamefont {Yan}, \citenamefont
  {Li},\ and\ \citenamefont {Shi}}]{yang2019magnetic}%
  \BibitemOpen
  \bibfield  {author} {\bibinfo {author} {\bibfnamefont {M.}~\bibnamefont
  {Yang}}, \bibinfo {author} {\bibfnamefont {G.}~\bibnamefont {Gu}}, \bibinfo
  {author} {\bibfnamefont {C.}~\bibnamefont {Yi}}, \bibinfo {author}
  {\bibfnamefont {D.}~\bibnamefont {Yan}}, \bibinfo {author} {\bibfnamefont
  {Y.}~\bibnamefont {Li}},\ and\ \bibinfo {author} {\bibfnamefont
  {Y.}~\bibnamefont {Shi}},\ }\bibfield  {title} {\bibinfo {title} {Magnetic
  and transport properties of Zr$_{1-x}$Nb$_x$Co$_2$Sn},\ }\href@noop {} {\bibfield
  {journal} {\bibinfo  {journal} {J. Phys.: Condens. Matter}\
  }\textbf {\bibinfo {volume} {31}},\ \bibinfo {pages} {275702} (\bibinfo
  {year} {2019})}\BibitemShut {NoStop}%
\bibitem [{\citenamefont {Barth}\ \emph {et~al.}(2010)\citenamefont {Barth},
  \citenamefont {Fecher}, \citenamefont {Balke}, \citenamefont {Ouardi},
  \citenamefont {Graf}, \citenamefont {Felser}, \citenamefont {Shkabko},
  \citenamefont {Weidenkaff}, \citenamefont {Klaer}, \citenamefont {Elmers}
  \emph {et~al.}}]{barth2010itinerant}%
  \BibitemOpen
  \bibfield  {author} {\bibinfo {author} {\bibfnamefont {J.}~\bibnamefont
  {Barth}}, \bibinfo {author} {\bibfnamefont {G.~H.}\ \bibnamefont {Fecher}},
  \bibinfo {author} {\bibfnamefont {B.}~\bibnamefont {Balke}}, \bibinfo
  {author} {\bibfnamefont {S.}~\bibnamefont {Ouardi}}, \bibinfo {author}
  {\bibfnamefont {T.}~\bibnamefont {Graf}}, \bibinfo {author} {\bibfnamefont
  {C.}~\bibnamefont {Felser}}, \bibinfo {author} {\bibfnamefont
  {A.}~\bibnamefont {Shkabko}}, \bibinfo {author} {\bibfnamefont
  {A.}~\bibnamefont {Weidenkaff}}, \bibinfo {author} {\bibfnamefont
  {P.}~\bibnamefont {Klaer}}, \bibinfo {author} {\bibfnamefont {H.~J.}\
  \bibnamefont {Elmers}}, \bibinfo {author} {\bibfnamefont {H.}\
  \bibnamefont {Yoshikawa}}, \bibinfo {author} {\bibfnamefont {S.}\
  \bibnamefont {Ueda}},\ and\ \bibinfo {author} {\bibfnamefont {K.}\
  \bibnamefont {Kobayashi}}, \ }\bibfield  {title} {\bibinfo {title} {Itinerant half-metallic ferromagnets Co$_2$TiZ (Z = Si, Ge, Sn): Ab
  initio calculations and measurement of the electronic structure and transport
  properties},\ }\href@noop {} {\bibfield  {journal} {\bibinfo  {journal}
  {Phys. Rev. B}\ }\textbf {\bibinfo {volume} {81}},\ \bibinfo {pages}
  {064404} (\bibinfo {year} {2010})}\BibitemShut {NoStop}%
\bibitem [{\citenamefont {Barth}\ \emph {et~al.}(2011)\citenamefont {Barth},
  \citenamefont {Fecher}, \citenamefont {Balke}, \citenamefont {Graf},
  \citenamefont {Shkabko}, \citenamefont {Weidenkaff}, \citenamefont {Klaer},
  \citenamefont {Kallmayer}, \citenamefont {Elmers}, \citenamefont {Yoshikawa}
  \emph {et~al.}}]{barth2011anomalous}%
  \BibitemOpen
  \bibfield  {author} {\bibinfo {author} {\bibfnamefont {J.}~\bibnamefont
  {Barth}}, \bibinfo {author} {\bibfnamefont {G.~H.}\ \bibnamefont {Fecher}},
  \bibinfo {author} {\bibfnamefont {B.}~\bibnamefont {Balke}}, \bibinfo
  {author} {\bibfnamefont {T.}~\bibnamefont {Graf}}, \bibinfo {author}
  {\bibfnamefont {A.}~\bibnamefont {Shkabko}}, \bibinfo {author} {\bibfnamefont
  {A.}~\bibnamefont {Weidenkaff}}, \bibinfo {author} {\bibfnamefont
  {P.}~\bibnamefont {Klaer}}, \bibinfo {author} {\bibfnamefont
  {M.}~\bibnamefont {Kallmayer}}, \bibinfo {author} {\bibfnamefont {H.-J.}\
  \bibnamefont {Elmers}}, \bibinfo {author} {\bibfnamefont {H.}~\bibnamefont
  {Yoshikawa}}, \bibinfo {author} {\bibfnamefont {S.}~\bibnamefont
  {Ueda}}, \bibinfo {author} {\bibfnamefont {K.}~\bibnamefont
  {Kobayashi}},\ and\ \bibinfo {author} {\bibfnamefont {C.}~\bibnamefont
  {Felser}}, \ }\bibfield  {title} {\bibinfo {title} {Anomalous transport properties of the half-metallic ferromagnets Co$_2$TiSi,
  Co$_2$TiGe and Co$_2$TiSn},\ }\href@noop {} {\bibfield  {journal} {\bibinfo
  {journal} {Phil. Trans. R. Soc. A.}\ }\textbf {\bibinfo {volume} {369}},\
  \bibinfo {pages} {3588} (\bibinfo {year} {2011})}\BibitemShut {NoStop}%
\bibitem [{\citenamefont {Ooka}\ \emph {et~al.}(2016)\citenamefont {Ooka},
  \citenamefont {Shigeta}, \citenamefont {Sukino}, \citenamefont {Fujimoto},
  \citenamefont {Umetsu}, \citenamefont {Miura}, \citenamefont {Nomura},
  \citenamefont {Yubuta}, \citenamefont {Yamauchi}, \citenamefont {Kanomata}
  \emph {et~al.}}]{ooka2016magnetization}%
  \BibitemOpen
  \bibfield  {author} {\bibinfo {author} {\bibfnamefont {R.}~\bibnamefont
  {Ooka}}, \bibinfo {author} {\bibfnamefont {I.}~\bibnamefont {Shigeta}},
  \bibinfo {author} {\bibfnamefont {Y.}~\bibnamefont {Sukino}}, \bibinfo
  {author} {\bibfnamefont {Y.}~\bibnamefont {Fujimoto}}, \bibinfo {author}
  {\bibfnamefont {R.~Y.}\ \bibnamefont {Umetsu}}, \bibinfo {author}
  {\bibfnamefont {Y.}~\bibnamefont {Miura}}, \bibinfo {author} {\bibfnamefont
  {A.}~\bibnamefont {Nomura}}, \bibinfo {author} {\bibfnamefont
  {K.}~\bibnamefont {Yubuta}}, \bibinfo {author} {\bibfnamefont
  {T.}~\bibnamefont {Yamauchi}}, \bibinfo {author} {\bibfnamefont
  {T.}~\bibnamefont {Kanomata}},\ and\ \bibinfo {author} {\bibfnamefont
  {M.}~\bibnamefont {Hiroi}}, \ }\bibfield  {title} {\bibinfo {title} {Magnetization and spin polarization of Heusler alloys Co$_2$TiSn and Co$_2$TiGa$_{0.5}$Sn$_{0.5}$},\ }\href@noop {} {\bibfield  {journal} {\bibinfo  {journal} {IEEE Magn. Lett.}\ }\textbf {\bibinfo {volume} {8}},\ \bibinfo {pages}
  {3101604} (\bibinfo {year} {2017})}\BibitemShut {NoStop}%
\bibitem [{\citenamefont {Bainsla}\ and\ \citenamefont
  {Suresh}(2016)}]{bainsla2016spin}%
  \BibitemOpen
  \bibfield  {author} {\bibinfo {author} {\bibfnamefont {L.}~\bibnamefont
  {Bainsla}}\ and\ \bibinfo {author} {\bibfnamefont {K. G.}~\bibnamefont
  {Suresh}},\ }\bibfield  {title} {\bibinfo {title} {Spin polarization studies
  in half-metallic Co$_2$TiX (X = Ge and Sn) Heusler alloys},\ }\href@noop {}
  {\bibfield  {journal} {\bibinfo  {journal} {Curr. Appl. Phys.}\
  }\textbf {\bibinfo {volume} {16}},\ \bibinfo {pages} {68} (\bibinfo {year}
  {2016})}\BibitemShut {NoStop}%
\bibitem [{\citenamefont {Shigeta}\ \emph {et~al.}(2018)\citenamefont
  {Shigeta}, \citenamefont {Fujimoto}, \citenamefont {Ooka}, \citenamefont
  {Nishisako}, \citenamefont {Tsujikawa}, \citenamefont {Umetsu}, \citenamefont
  {Nomura}, \citenamefont {Yubuta}, \citenamefont {Miura}, \citenamefont
  {Kanomata} \emph {et~al.}}]{shigeta2018pressure}%
  \BibitemOpen
  \bibfield  {author} {\bibinfo {author} {\bibfnamefont {I.}~\bibnamefont
  {Shigeta}}, \bibinfo {author} {\bibfnamefont {Y.}~\bibnamefont {Fujimoto}},
  \bibinfo {author} {\bibfnamefont {R.}~\bibnamefont {Ooka}}, \bibinfo {author}
  {\bibfnamefont {Y.}~\bibnamefont {Nishisako}}, \bibinfo {author}
  {\bibfnamefont {M.}~\bibnamefont {Tsujikawa}}, \bibinfo {author}
  {\bibfnamefont {R.~Y.}\ \bibnamefont {Umetsu}}, \bibinfo {author}
  {\bibfnamefont {A.}~\bibnamefont {Nomura}}, \bibinfo {author} {\bibfnamefont
  {K.}~\bibnamefont {Yubuta}}, \bibinfo {author} {\bibfnamefont
  {Y.}~\bibnamefont {Miura}}, \bibinfo {author} {\bibfnamefont
  {T.}~\bibnamefont {Kanomata}}, \bibinfo {author} {\bibfnamefont
  {M.}~\bibnamefont {Shirai}}, \bibinfo {author} {\bibfnamefont
  {J.}~\bibnamefont {Gouchi}}, \bibinfo {author} {\bibfnamefont
  {Y.}~\bibnamefont {Uwatoko}},\ and\ \bibinfo {author} {\bibfnamefont
  {T.}~\bibnamefont {Hiroi}}, \ }\bibfield  {title} {\bibinfo {title} {Pressure effect on the magnetic properties of the half-metallic
  Heusler alloy Co$_2$TiSn},\ }\href@noop {} {\bibfield  {journal} {\bibinfo
  {journal} {Phys. Rev. B}\ }\textbf {\bibinfo {volume} {97}},\ \bibinfo
  {pages} {104414} (\bibinfo {year} {2018})}\BibitemShut {NoStop}%
\bibitem [{\citenamefont {Hu}\ \emph {et~al.}(2019)\citenamefont {Hu},
  \citenamefont {Niu}, \citenamefont {Ernst}, \citenamefont {Tu}, \citenamefont
  {Hamzić}, \citenamefont {Liu}, \citenamefont {Zhang}, \citenamefont {Wu},
  \citenamefont {Felser},\ and\ \citenamefont {Yu}}]{hu2019unconventional}%
  \BibitemOpen
  \bibfield  {author} {\bibinfo {author} {\bibfnamefont {J.}~\bibnamefont
  {Hu}}, \bibinfo {author} {\bibfnamefont {J.}~\bibnamefont {Niu}}, \bibinfo
  {author} {\bibfnamefont {B.}~\bibnamefont {Ernst}}, \bibinfo {author}
  {\bibfnamefont {S.}~\bibnamefont {Tu}}, \bibinfo {author} {\bibfnamefont
  {A.}~\bibnamefont {Hamzić}}, \bibinfo {author} {\bibfnamefont
  {C.}~\bibnamefont {Liu}}, \bibinfo {author} {\bibfnamefont {Y.}~\bibnamefont
  {Zhang}}, \bibinfo {author} {\bibfnamefont {X.}~\bibnamefont {Wu}}, \bibinfo
  {author} {\bibfnamefont {C.}~\bibnamefont {Felser}},\ and\ \bibinfo {author}
  {\bibfnamefont {H.}~\bibnamefont {Yu}},\ }\bibfield  {title} {\bibinfo
  {title} {Unconventional spin-dependent thermopower in epitaxial
  Co$_2$Ti$_{0.6}$V$_{0.4}$Sn$_{0.75}$ Heusler film},\ }\href@noop {} {\bibfield  {journal} {\bibinfo  {journal} {Solid State Commun.}\ }\textbf {\bibinfo
  {volume} {299}},\ \bibinfo {pages} {113661} (\bibinfo {year}
  {2019})}\BibitemShut {NoStop}%
\bibitem [{\citenamefont {Hu}\ \emph {et~al.}(2018)\citenamefont {Hu},
  \citenamefont {Ernst}, \citenamefont {Tu}, \citenamefont
  {Kuve{\v{z}}di{\'c}}, \citenamefont {Hamzi{\'c}}, \citenamefont {Tafra},
  \citenamefont {Basleti{\'c}}, \citenamefont {Zhang}, \citenamefont {Markou},
  \citenamefont {Felser} \emph {et~al.}}]{hu2018anomalous}%
  \BibitemOpen
  \bibfield  {author} {\bibinfo {author} {\bibfnamefont {J.}~\bibnamefont
  {Hu}}, \bibinfo {author} {\bibfnamefont {B.}~\bibnamefont {Ernst}}, \bibinfo
  {author} {\bibfnamefont {S.}~\bibnamefont {Tu}}, \bibinfo {author}
  {\bibfnamefont {M.}~\bibnamefont {Kuve{\v{z}}di{\'c}}}, \bibinfo {author}
  {\bibfnamefont {A.}~\bibnamefont {Hamzi{\'c}}}, \bibinfo {author}
  {\bibfnamefont {E.}~\bibnamefont {Tafra}}, \bibinfo {author} {\bibfnamefont
  {M.}~\bibnamefont {Basleti{\'c}}}, \bibinfo {author} {\bibfnamefont
  {Y.}~\bibnamefont {Zhang}}, \bibinfo {author} {\bibfnamefont
  {A.}~\bibnamefont {Markou}}, \bibinfo {author} {\bibfnamefont
  {C.}~\bibnamefont {Felser}}, \bibinfo {author} {\bibfnamefont
  {A.}~\bibnamefont {Fert}}, \bibinfo {author} {\bibfnamefont
  {W.}~\bibnamefont {Zhao}}, \bibinfo {author} {\bibfnamefont
  {J.}~\bibnamefont {Ansermet}},\ and\ \bibinfo {author} {\bibfnamefont
  {H.}~\bibnamefont {Hu}}, \ }\bibfield  {title} {\bibinfo {title} {Anomalous Hall and Nernst effects in Co$_2$TiSn and Co$_2$Ti$_{0.6}$V$_{0.4}$Sn Heusler thin films},\ }\href@noop {} {\bibfield  {journal} {\bibinfo
  {journal} {Phys. Rev. Appl.}\ }\textbf {\bibinfo {volume} {10}},\
  \bibinfo {pages} {044037} (\bibinfo {year} {2018})}\BibitemShut {NoStop}%
\bibitem [{\citenamefont {Dunlap}\ and\ \citenamefont
  {Stroink}(1982)}]{dunlap1982conduction}%
  \BibitemOpen
  \bibfield  {author} {\bibinfo {author} {\bibfnamefont {R.}~\bibnamefont
  {Dunlap}}\ and\ \bibinfo {author} {\bibfnamefont {G.}~\bibnamefont
  {Stroink}},\ }\bibfield  {title} {\bibinfo {title} {Conduction electron
  contributions to the Sn hyperfine field in the Heusler alloy Co$_2$Ti$_{1-x}$V$_x$Sn},\
  }\href@noop {} {\bibfield  {journal} {\bibinfo  {journal} {J. Appl. Phys.}\ }\textbf {\bibinfo {volume} {53}},\ \bibinfo {pages} {8210}
  (\bibinfo {year} {1982})}\BibitemShut {NoStop}%
\bibitem [{\citenamefont {Pendl~Jr}\ \emph {et~al.}(1996)\citenamefont
  {Pendl~Jr}, \citenamefont {Saxena}, \citenamefont {Carbonari}, \citenamefont
  {Mestnik~Filho},\ and\ \citenamefont {Schaff}}]{pendl1996investigation}%
  \BibitemOpen
  \bibfield  {author} {\bibinfo {author} {\bibfnamefont {W.}~\bibnamefont
  {Pendl~Jr}}, \bibinfo {author} {\bibfnamefont {R.}~\bibnamefont {Saxena}},
  \bibinfo {author} {\bibfnamefont {A. W.}~\bibnamefont {Carbonari}}, \bibinfo
  {author} {\bibfnamefont {J.}~\bibnamefont {Mestnik~Filho}},\ and\ \bibinfo
  {author} {\bibfnamefont {J.}~\bibnamefont {Schaff}},\ }\bibfield  {title}
  {\bibinfo {title} {Investigation of the magnetic hyperfine field at the Y
  site in the Heusler alloys (Y = Ti, V, Nb, Cr; Z = Al, Sn)},\ }\href@noop {}
  {\bibfield  {journal} {\bibinfo  {journal} {J. Phys.: Condens. Matter}\ }\textbf {\bibinfo {volume} {8}},\ \bibinfo {pages} {11317}
  (\bibinfo {year} {1996})}\BibitemShut {NoStop}%
\bibitem [{\citenamefont {Ebert}\ \emph {et~al.}(2005)\citenamefont {Ebert} \emph {et~al.}}]{ebert2005munich}%
   \BibitemOpen
  \bibfield  {author} {\bibinfo {author} {\bibfnamefont {H.}~\bibnamefont 
  {Ebert}}, \bibinfo {author} {\bibfnamefont {M.}~\bibnamefont {Battocletti}},
  \bibinfo {author} {\bibfnamefont {D.}~\bibnamefont {Benea}}, \bibinfo
  {author} {\bibfnamefont {S.}~\bibnamefont {Bornemann}}, \bibinfo {author}
  {\bibfnamefont {J.}~\bibnamefont {Braun}}, \bibinfo {author} {\bibfnamefont
  {S.}\ \bibnamefont {Chadov}}, \bibinfo {author} {\bibfnamefont
  {M.}~\bibnamefont {Deng}}, \bibinfo {author} {\bibfnamefont
  {H.}~\bibnamefont {Freyer}}, \bibinfo {author} {\bibfnamefont
  {T.}~\bibnamefont {Huhne}}, \bibinfo {author} {\bibfnamefont
  {D.}~\bibnamefont {Kodderitzsch}}, \bibinfo {author} {\bibfnamefont
  {M.}~\bibnamefont {Kosuth}}, \bibinfo {author} {\bibfnamefont
  {S.}~\bibnamefont {Lowitzer}}, \bibinfo {author} {\bibfnamefont
  {S.}~\bibnamefont {Mankovskyy}}, \bibinfo {author} {\bibfnamefont
  {J.}~\bibnamefont {Minar}}, \bibinfo {author} {\bibfnamefont
  {A.}~\bibnamefont {Perlov}}, \bibinfo {author} {\bibfnamefont {V.}~\bibnamefont {Popescu}},\bibinfo {author} {\bibfnamefont
  {O.}~\bibnamefont {Sipr}}, \bibinfo {author} {\bibfnamefont
  {Ch.}~\bibnamefont {Zecha}}, \ } \bibfield  {title} {\bibinfo {title} {The munich SPRKKR package, version 6.3},\ }\href@noop {} {\bibfield  {journal} {\bibinfo  {journal} {URL:https://www.ebert.cup.uni-muenchen.de/index.php/de/.}\ }
  (\bibinfo {year} {2012})}\BibitemShut {NoStop}%
\bibitem [{\citenamefont {Perdew}\ \emph {et~al.}(1996)\citenamefont {Perdew},
  \citenamefont {Burke},\ and\ \citenamefont
  {Ernzerhof}}]{perdew1996generalized}%
  \BibitemOpen
  \bibfield  {author} {\bibinfo {author} {\bibfnamefont {J.~P.}\ \bibnamefont
  {Perdew}}, \bibinfo {author} {\bibfnamefont {K.}~\bibnamefont {Burke}},\ and\
  \bibinfo {author} {\bibfnamefont {M.}~\bibnamefont {Ernzerhof}},\ }\bibfield
  {title} {\bibinfo {title} {Generalized gradient approximation made simple},\
  }\href@noop {} {\bibfield  {journal} {\bibinfo  {journal} {Phys. Rev. Lett.}\ }\textbf {\bibinfo {volume} {77}},\ \bibinfo {pages} {3865}
  (\bibinfo {year} {1996})}\BibitemShut {NoStop}%
\bibitem [{\citenamefont {Ebert}\ \emph {et~al.}(2011)\citenamefont {Ebert},
  \citenamefont {Koedderitzsch},\ and\ \citenamefont
  {Minar}}]{ebert2011calculating}%
  \BibitemOpen
  \bibfield  {author} {\bibinfo {author} {\bibfnamefont {H.}~\bibnamefont
  {Ebert}}, \bibinfo {author} {\bibfnamefont {D.}~\bibnamefont
  {Koedderitzsch}},\ and\ \bibinfo {author} {\bibfnamefont {J.}~\bibnamefont
  {Minar}},\ }\bibfield  {title} {\bibinfo {title} {Calculating condensed
  matter properties using the KKR-Green's function method—recent developments
  and applications},\ }\href@noop {} {\bibfield  {journal} {\bibinfo  {journal}
  {Rep. Prog. Phys.}\ }\textbf {\bibinfo {volume} {74}},\
  \bibinfo {pages} {096501} (\bibinfo {year} {2011})}\BibitemShut {NoStop}%
\bibitem [{\citenamefont {Soven}(1967)}]{soven1967coherent}%
  \BibitemOpen
  \bibfield  {author} {\bibinfo {author} {\bibfnamefont {P.}~\bibnamefont
  {Soven}},\ }\bibfield  {title} {\bibinfo {title} {Coherent-potential model of
  substitutional disordered alloys},\ }\href@noop {} {\bibfield  {journal}
  {\bibinfo  {journal} {Phys. Rev.}\ }\textbf {\bibinfo {volume} {156}},\
  \bibinfo {pages} {809} (\bibinfo {year} {1967})}\BibitemShut {NoStop}%
\bibitem [{\citenamefont {Soven}(1969)}]{soven1969contribution}%
  \BibitemOpen
  \bibfield  {author} {\bibinfo {author} {\bibfnamefont {P.}~\bibnamefont
  {Soven}},\ }\bibfield  {title} {\bibinfo {title} {Contribution to the theory
  of disordered alloys},\ }\href@noop {} {\bibfield  {journal} {\bibinfo
  {journal} {Phys. Rev.}\ }\textbf {\bibinfo {volume} {178}},\ \bibinfo
  {pages} {1136} (\bibinfo {year} {1969})}\BibitemShut {NoStop}%
\bibitem [{\citenamefont {Giannozzi}\ \emph {et~al.}(2009)\citenamefont
  {Giannozzi}, \citenamefont {Baroni}, \citenamefont {Bonini}, \citenamefont
  {Calandra}, \citenamefont {Car}, \citenamefont {Cavazzoni}, \citenamefont
  {Ceresoli}, \citenamefont {Chiarotti}, \citenamefont {Cococcioni},
  \citenamefont {Dabo} \emph {et~al.}}]{giannozzi2009quantum}%
  \BibitemOpen
  \bibfield  {author} {\bibinfo {author} {\bibfnamefont {P.}~\bibnamefont
  {Giannozzi}}, \bibinfo {author} {\bibfnamefont {S.}~\bibnamefont {Baroni}},
  \bibinfo {author} {\bibfnamefont {N.}~\bibnamefont {Bonini}}, \bibinfo
  {author} {\bibfnamefont {M.}~\bibnamefont {Calandra}}, \bibinfo {author}
  {\bibfnamefont {R.}~\bibnamefont {Car}}, \bibinfo {author} {\bibfnamefont
  {C.}~\bibnamefont {Cavazzoni}}, \bibinfo {author} {\bibfnamefont
  {D.}~\bibnamefont {Ceresoli}}, \bibinfo {author} {\bibfnamefont {G.~L.}\
  \bibnamefont {Chiarotti}}, \bibinfo {author} {\bibfnamefont {M.}~\bibnamefont
  {Cococcioni}}, \bibinfo {author} {\bibfnamefont {I.}~\bibnamefont {Dabo}}, \bibinfo {author} {\bibfnamefont {A. D.}~\bibnamefont {Corso}},  \bibinfo {author} {\bibfnamefont {S. d.}~\bibnamefont {Gironcoli}},  \bibinfo {author} {\bibfnamefont {S.}~\bibnamefont {Fabris}},  \bibinfo {author} {\bibfnamefont {G.}~\bibnamefont {Fratesi}},  \bibinfo {author} {\bibfnamefont {R.}~\bibnamefont {Gebauer}},  \bibinfo {author} {\bibfnamefont {U.}~\bibnamefont {Gerstmann}},  \bibinfo {author} {\bibfnamefont {C.}~\bibnamefont {Gougoussis}},  \bibinfo {author} {\bibfnamefont {A.}~\bibnamefont { Kokalj}},  \bibinfo {author} {\bibfnamefont {M.}~\bibnamefont {Lazzeri}},  \bibinfo {author} {\bibfnamefont {L.}~\bibnamefont {Martin-Samos}},  \bibinfo {author} {\bibfnamefont {N.}~\bibnamefont {Marzari}},  \bibinfo {author} {\bibfnamefont {F.}~\bibnamefont {Mauri}},  \bibinfo {author} {\bibfnamefont {R.}~\bibnamefont {Mazzarello}},  \bibinfo {author} {\bibfnamefont {S.}~\bibnamefont {Paolini}},  \bibinfo {author} {\bibfnamefont {A.}~\bibnamefont {Pasquarello}},  \bibinfo {author} {\bibfnamefont {L.}~\bibnamefont {Paulatto}},  \bibinfo {author} {\bibfnamefont {C.}~\bibnamefont {Sbraccia}},  \bibinfo {author} {\bibfnamefont {S.}~\bibnamefont {Scandolo}},  \bibinfo {author} {\bibfnamefont {G.}~\bibnamefont {Sclauzero}},  \bibinfo {author} {\bibfnamefont {A. P.}~\bibnamefont {Seitsonen}},  \bibinfo {author} {\bibfnamefont {A.}~\bibnamefont {Smogunov}},  \bibinfo {author} {\bibfnamefont {P.}~\bibnamefont {Umari}},\ and\  \bibinfo {author} {\bibfnamefont {R. W.}~\bibnamefont {Wentzcovitch}}, \ }\bibfield  {title} {\bibinfo {title} {Quantum Espresso: a
  modular and open-source software project for quantum simulations of
  materials},\ }\href@noop {} {\bibfield  {journal} {\bibinfo  {journal}
  {J. Phys. Condens. Matter}\ }\textbf {\bibinfo {volume} {21}},\
  \bibinfo {pages} {395502} (\bibinfo {year} {2009})}\BibitemShut {NoStop}%
\bibitem [{\citenamefont {Kresse}\ and\ \citenamefont
  {Hafner}(1993)}]{kresse1993ab}%
  \BibitemOpen
  \bibfield  {author} {\bibinfo {author} {\bibfnamefont {G.}~\bibnamefont
  {Kresse}}\ and\ \bibinfo {author} {\bibfnamefont {J.}~\bibnamefont
  {Hafner}},\ }\bibfield  {title} {\bibinfo {title} {Ab initio molecular
  dynamics for liquid metals},\ }\href@noop {} {\bibfield  {journal} {\bibinfo
  {journal} {Phys. Rev. B}\ }\textbf {\bibinfo {volume} {47}},\ \bibinfo
  {pages} {558} (\bibinfo {year} {1993})}\BibitemShut {NoStop}%
\bibitem [{\citenamefont {Kresse}\ and\ \citenamefont
  {Furthm{\"u}ller}(1996)}]{kresse1996efficient}%
  \BibitemOpen
  \bibfield  {author} {\bibinfo {author} {\bibfnamefont {G.}~\bibnamefont
  {Kresse}}\ and\ \bibinfo {author} {\bibfnamefont {J.}~\bibnamefont
  {Furthm{\"u}ller}},\ }\bibfield  {title} {\bibinfo {title} {Efficient
  iterative schemes for ab initio total-energy calculations using a plane-wave
  basis set},\ }\href@noop {} {\bibfield  {journal} {\bibinfo  {journal}
  {Phys. Rev. B}\ }\textbf {\bibinfo {volume} {54}},\ \bibinfo {pages}
  {11169} (\bibinfo {year} {1996})}\BibitemShut {NoStop}%
\bibitem [{\citenamefont {Marzari}\ and\ \citenamefont
  {Vanderbilt}(1997)}]{marzari1997maximally}%
  \BibitemOpen
  \bibfield  {author} {\bibinfo {author} {\bibfnamefont {N.}~\bibnamefont
  {Marzari}}\ and\ \bibinfo {author} {\bibfnamefont {D.}~\bibnamefont
  {Vanderbilt}},\ }\bibfield  {title} {\bibinfo {title} {Maximally localized
  generalized Wannier functions for composite energy bands},\ }\href@noop {}
  {\bibfield  {journal} {\bibinfo  {journal} {Phys. Rev. B}\ }\textbf
  {\bibinfo {volume} {56}},\ \bibinfo {pages} {12847} (\bibinfo {year}
  {1997})}\BibitemShut {NoStop}%
\bibitem [{\citenamefont {Souza}\ \emph {et~al.}(2001)\citenamefont {Souza},
  \citenamefont {Marzari},\ and\ \citenamefont
  {Vanderbilt}}]{souza2001maximally}%
  \BibitemOpen
  \bibfield  {author} {\bibinfo {author} {\bibfnamefont {I.}~\bibnamefont
  {Souza}}, \bibinfo {author} {\bibfnamefont {N.}~\bibnamefont {Marzari}},\
  and\ \bibinfo {author} {\bibfnamefont {D.}~\bibnamefont {Vanderbilt}},\
  }\bibfield  {title} {\bibinfo {title} {Maximally localized Wannier functions
  for entangled energy bands},\ }\href@noop {} {\bibfield  {journal} {\bibinfo
  {journal} {Phys. Rev. B}\ }\textbf {\bibinfo {volume} {65}},\ \bibinfo
  {pages} {035109} (\bibinfo {year} {2001})}\BibitemShut {NoStop}%
\bibitem [{\citenamefont {Pizzi}\ \emph {et~al.}(2020)\citenamefont {Pizzi},
  \citenamefont {Vitale}, \citenamefont {Arita}, \citenamefont {Bl{\"u}gel},
  \citenamefont {Freimuth}, \citenamefont {G{\'e}ranton}, \citenamefont
  {Gibertini}, \citenamefont {Gresch}, \citenamefont {Johnson}, \citenamefont
  {Koretsune} \emph {et~al.}}]{pizzi2020wannier90}%
  \BibitemOpen
  \bibfield  {author} {\bibinfo {author} {\bibfnamefont {G.}~\bibnamefont
  {Pizzi}}, \bibinfo {author} {\bibfnamefont {V.}~\bibnamefont {Vitale}},
  \bibinfo {author} {\bibfnamefont {R.}~\bibnamefont {Arita}}, \bibinfo
  {author} {\bibfnamefont {S.}~\bibnamefont {Bl{\"u}gel}}, \bibinfo {author}
  {\bibfnamefont {F.}~\bibnamefont {Freimuth}}, \bibinfo {author}
  {\bibfnamefont {G.}~\bibnamefont {G{\'e}ranton}}, \bibinfo {author}
  {\bibfnamefont {M.}~\bibnamefont {Gibertini}}, \bibinfo {author}
  {\bibfnamefont {D.}~\bibnamefont {Gresch}}, \bibinfo {author} {\bibfnamefont
  {C.}~\bibnamefont {Johnson}}, \bibinfo {author} {\bibfnamefont
  {T.}~\bibnamefont {Koretsune}}, \bibinfo {author} {\bibfnamefont
  {J.}~\bibnamefont {Ibañez-Azpiroz}}, \bibinfo {author} {\bibfnamefont
  {H.}~\bibnamefont {Lee}}, \bibinfo {author} {\bibfnamefont
  {J.$-$M.}~\bibnamefont {Lihm}}, \bibinfo {author} {\bibfnamefont
  {D.}~\bibnamefont {Marchand}}, \bibinfo {author} {\bibfnamefont
  {A.}~\bibnamefont {Marrazzo}}, \bibinfo {author} {\bibfnamefont
  {Y.}~\bibnamefont {Mokrousov}}, \bibinfo {author} {\bibfnamefont
  {J. I.}~\bibnamefont {Mustafa}}, \bibinfo {author} {\bibfnamefont
  {Y.}~\bibnamefont {Nohara}}, \bibinfo {author} {\bibfnamefont
  {Y.}~\bibnamefont {Nomura}}, \bibinfo {author} {\bibfnamefont
  {L.}~\bibnamefont {Paulatto}}, \bibinfo {author} {\bibfnamefont
  {S.}~\bibnamefont {Poncé}}, \bibinfo {author} {\bibfnamefont
  {T.}~\bibnamefont {Ponweiser}}, \bibinfo {author} {\bibfnamefont
  {J.}~\bibnamefont {Qiao}}, \bibinfo {author} {\bibfnamefont
  {F.}~\bibnamefont {Thöle}}, \bibinfo {author} {\bibfnamefont
  {S. S.}~\bibnamefont {Tsirkin}}, \bibinfo {author} {\bibfnamefont
  {M.}~\bibnamefont {Wierzbowska}}, \bibinfo {author} {\bibfnamefont
  {N.}~\bibnamefont {Marzari}}, \bibinfo {author} {\bibfnamefont
  {D.}~\bibnamefont {Vanderbilt}}, \bibinfo {author} {\bibfnamefont
  {I.}~\bibnamefont {Souza}}, \bibinfo {author} {\bibfnamefont
  {A. A.}~\bibnamefont {Mostofi}},\ and\ \bibinfo {author} {\bibfnamefont
  {J. R.}~\bibnamefont {Yates}},\ }\bibfield  {title}
  {\bibinfo {title} {Wannier90 as a community code: new features and
  applications},\ }\href@noop {} {\bibfield  {journal} {\bibinfo  {journal}
  {J. Phys. Condens. Matter}\ }\textbf {\bibinfo {volume} {32}},\
  \bibinfo {pages} {165902} (\bibinfo {year} {2020})}\BibitemShut {NoStop}%
\bibitem [{\citenamefont {Wu}\ \emph {et~al.}(2018)\citenamefont {Wu},
  \citenamefont {Zhang}, \citenamefont {Song}, \citenamefont {Troyer},\ and\
  \citenamefont {Soluyanov}}]{wu2018wanniertools}%
  \BibitemOpen
  \bibfield  {author} {\bibinfo {author} {\bibfnamefont {Q.}~\bibnamefont
  {Wu}}, \bibinfo {author} {\bibfnamefont {S.}~\bibnamefont {Zhang}}, \bibinfo
  {author} {\bibfnamefont {H.-F.}\ \bibnamefont {Song}}, \bibinfo {author}
  {\bibfnamefont {M.}~\bibnamefont {Troyer}},\ and\ \bibinfo {author}
  {\bibfnamefont {A.~A.}\ \bibnamefont {Soluyanov}},\ }\bibfield  {title}
  {\bibinfo {title} {WannierTools: An open-source software package for novel
  topological materials},\ }\href@noop {} {\bibfield  {journal} {\bibinfo
  {journal} {Comput. Phys. Commun.}\ }\textbf {\bibinfo {volume}
  {224}},\ \bibinfo {pages} {405} (\bibinfo {year} {2018})}\BibitemShut
  {NoStop}%
\bibitem [{\citenamefont {Hamann}(2013)}]{hamann2013optimized}%
  \BibitemOpen
  \bibfield  {author} {\bibinfo {author} {\bibfnamefont {D.}~\bibnamefont
  {Hamann}},\ }\bibfield  {title} {\bibinfo {title} {Optimized norm-conserving
  vanderbilt pseudopotentials},\ }\href@noop {} {\bibfield  {journal} {\bibinfo
   {journal} {Phys. Rev. B}\ }\textbf {\bibinfo {volume} {88}},\ \bibinfo
  {pages} {085117} (\bibinfo {year} {2013})}\BibitemShut {NoStop}%
\bibitem [{\citenamefont {Birch}(1947)}]{birch1947finite}%
  \BibitemOpen
  \bibfield  {author} {\bibinfo {author} {\bibfnamefont {F.}~\bibnamefont
  {Birch}},\ }\bibfield  {title} {\bibinfo {title} {Finite elastic strain of
  cubic crystals},\ }\href@noop {} {\bibfield  {journal} {\bibinfo  {journal}
  {Phys. Rev.}\ }\textbf {\bibinfo {volume} {71}},\ \bibinfo {pages} {809}
  (\bibinfo {year} {1947})}\BibitemShut {NoStop}%
\bibitem [{\citenamefont {Van~Engen}\ \emph {et~al.}(1983)\citenamefont
  {Van~Engen}, \citenamefont {Buschow},\ and\ \citenamefont
  {Erman}}]{van1983magnetic}%
  \BibitemOpen
  \bibfield  {author} {\bibinfo {author} {\bibfnamefont {P. G.}~\bibnamefont
  {Van~Engen}}, \bibinfo {author} {\bibfnamefont {K. H. J.}~\bibnamefont {Buschow}},\
  and\ \bibinfo {author} {\bibfnamefont {M.}~\bibnamefont {Erman}},\ }\bibfield
   {title} {\bibinfo {title} {Magnetic properties and magneto-optical
  spectroscopy of Heusler alloys based on transition metals and Sn},\
  }\href@noop {} {\bibfield  {journal} {\bibinfo  {journal} {J. Magn. Magn. Mater.}\ }\textbf {\bibinfo {volume} {30}},\
  \bibinfo {pages} {374} (\bibinfo {year} {1983})}\BibitemShut {NoStop}%
\bibitem [{\citenamefont {Carbonari}\ \emph {et~al.}(1996)\citenamefont
  {Carbonari}, \citenamefont {Saxena}, \citenamefont {Pendl~Jr}, \citenamefont
  {Mestnik~Filho}, \citenamefont {Attili}, \citenamefont {Olzon-Dionysio},\
  and\ \citenamefont {De~Souza}}]{carbonari1996magnetic}%
  \BibitemOpen
  \bibfield  {author} {\bibinfo {author} {\bibfnamefont {A. W.}~\bibnamefont
  {Carbonari}}, \bibinfo {author} {\bibfnamefont {R.}~\bibnamefont {Saxena}},
  \bibinfo {author} {\bibfnamefont {W.}~\bibnamefont {Pendl~Jr}}, \bibinfo
  {author} {\bibfnamefont {J.}~\bibnamefont {Mestnik~Filho}}, \bibinfo {author}
  {\bibfnamefont {R. N.}~\bibnamefont {Attili}}, \bibinfo {author} {\bibfnamefont
  {M.}~\bibnamefont {Olzon-Dionysio}},\ and\ \bibinfo {author} {\bibfnamefont
  {S. D.}~\bibnamefont {De~Souza}},\ }\bibfield  {title} {\bibinfo {title}
  {Magnetic hyperfine field in the Heusler alloys Co$_2$YZ (Y = V, Nb, Ta, Cr; Z =
  Al, Ga)},\ }\href@noop {} {\bibfield  {journal} {\bibinfo  {journal} {J. Magn. Magn. Mater.}\ }\textbf {\bibinfo {volume} {163}},\
  \bibinfo {pages} {313} (\bibinfo {year} {1996})}\BibitemShut {NoStop}%
\bibitem [{\citenamefont {Popescu}\ and\ \citenamefont
  {Zunger}(2010)}]{popescu2010effective}%
  \BibitemOpen
  \bibfield  {author} {\bibinfo {author} {\bibfnamefont {V.}~\bibnamefont
  {Popescu}}\ and\ \bibinfo {author} {\bibfnamefont {A.}~\bibnamefont
  {Zunger}},\ }\bibfield  {title} {\bibinfo {title} {Effective band structure
  of random alloys},\ }\href@noop {} {\bibfield  {journal} {\bibinfo  {journal}
  {Phys. Rev. Lett.}\ }\textbf {\bibinfo {volume} {104}},\ \bibinfo
  {pages} {236403} (\bibinfo {year} {2010})}\BibitemShut {NoStop}%
\bibitem [{\citenamefont {Popescu}\ and\ \citenamefont
  {Zunger}(2012)}]{popescu2012extracting}%
  \BibitemOpen
  \bibfield  {author} {\bibinfo {author} {\bibfnamefont {V.}~\bibnamefont
  {Popescu}}\ and\ \bibinfo {author} {\bibfnamefont {A.}~\bibnamefont
  {Zunger}},\ }\bibfield  {title} {\bibinfo {title} {Extracting E versus $\vec{k}$
  effective band structure from supercell calculations on alloys and
  impurities},\ }\href@noop {} {\bibfield  {journal} {\bibinfo  {journal}
  {Phys. Rev. B}\ }\textbf {\bibinfo {volume} {85}},\ \bibinfo {pages}
  {085201} (\bibinfo {year} {2012})}\BibitemShut {NoStop}%
\bibitem [{\citenamefont {Lee}\ \emph {et~al.}(2005)\citenamefont {Lee},
  \citenamefont {Lee}, \citenamefont {Blaha},\ and\ \citenamefont
  {Schwarz}}]{doi:10.1063/1.1853899}%
  \BibitemOpen
  \bibfield  {author} {\bibinfo {author} {\bibfnamefont {S.~C.}\ \bibnamefont
  {Lee}}, \bibinfo {author} {\bibfnamefont {T.~D.}\ \bibnamefont {Lee}},
  \bibinfo {author} {\bibfnamefont {P.}~\bibnamefont {Blaha}},\ and\ \bibinfo
  {author} {\bibfnamefont {K.}~\bibnamefont {Schwarz}},\ }\bibfield  {title}
  {\bibinfo {title} {Magnetic and half-metallic properties of the full-Heusler
  alloys Co$_2$TiX (X = Al, Ga; Si, Ge, Sn; Sb)},\ }\href@noop {} {\bibfield  {journal}
  {\bibinfo  {journal} {J. Appl. Phys.}\ }\textbf {\bibinfo
  {volume} {97}},\ \bibinfo {pages} {10C307} (\bibinfo {year}
  {2005})}\BibitemShut {NoStop}%
\bibitem [{\citenamefont {Hickey}\ \emph {et~al.}(2006)\citenamefont {Hickey},
  \citenamefont {Husmann}, \citenamefont {Holmes},\ and\ \citenamefont
  {Jones}}]{hickey2006Fermi}%
  \BibitemOpen
  \bibfield  {author} {\bibinfo {author} {\bibfnamefont {M.}~\bibnamefont
  {Hickey}}, \bibinfo {author} {\bibfnamefont {A.}~\bibnamefont {Husmann}},
  \bibinfo {author} {\bibfnamefont {S.}~\bibnamefont {Holmes}},\ and\ \bibinfo
  {author} {\bibfnamefont {G.}~\bibnamefont {Jones}},\ }\bibfield  {title}
  {\bibinfo {title} {Fermi surfaces and electronic structure of the Heusler
  alloy Co$_2$TiSn},\ }\href@noop {} {\bibfield  {journal} {\bibinfo  {journal}
  {J. Phys.: Condens. Matter}\ }\textbf {\bibinfo {volume} {18}},\
  \bibinfo {pages} {2897} (\bibinfo {year} {2006})}\BibitemShut {NoStop}%
\bibitem [{\citenamefont {Kandpal}\ \emph {et~al.}(2007)\citenamefont
  {Kandpal}, \citenamefont {Fecher},\ and\ \citenamefont
  {Felser}}]{kandpal2007calculated}%
  \BibitemOpen
  \bibfield  {author} {\bibinfo {author} {\bibfnamefont {H.~C.}\ \bibnamefont
  {Kandpal}}, \bibinfo {author} {\bibfnamefont {G.~H.}\ \bibnamefont
  {Fecher}},\ and\ \bibinfo {author} {\bibfnamefont {C.}~\bibnamefont
  {Felser}},\ }\bibfield  {title} {\bibinfo {title} {Calculated electronic and
  magnetic properties of the half-metallic, transition metal based Heusler
  compounds},\ }\href@noop {} {\bibfield  {journal} {\bibinfo  {journal}
  {J. Phys. D: Appl. Phys.}\ }\textbf {\bibinfo {volume} {40}},\
  \bibinfo {pages} {1507} (\bibinfo {year} {2007})}\BibitemShut {NoStop}%
\bibitem [{\citenamefont {Aguayo}\ and\ \citenamefont
  {Murrieta}(2011)}]{aguayo2011density}%
  \BibitemOpen
  \bibfield  {author} {\bibinfo {author} {\bibfnamefont {A.}~\bibnamefont
  {Aguayo}}\ and\ \bibinfo {author} {\bibfnamefont {G.}~\bibnamefont
  {Murrieta}},\ }\bibfield  {title} {\bibinfo {title} {Density functional study
  of the half-metallic ferromagnetism in Co-based Heusler alloys Co$_2$MSn (M = Ti,
  Zr, Hf) using LSDA and GGA},\ }\href@noop {} {\bibfield  {journal} {\bibinfo
  {journal} {J. Magn. Magn. Mater.}\ }\textbf {\bibinfo
  {volume} {323}},\ \bibinfo {pages} {3013} (\bibinfo {year}
  {2011})}\BibitemShut {NoStop}%
\bibitem [{\citenamefont {K{\"u}bler}\ and\ \citenamefont
  {Felser}(2012)}]{kubler2012berry}%
  \BibitemOpen
  \bibfield  {author} {\bibinfo {author} {\bibfnamefont {J.}~\bibnamefont
  {K{\"u}bler}}\ and\ \bibinfo {author} {\bibfnamefont {C.}~\bibnamefont
  {Felser}},\ }\bibfield  {title} {\bibinfo {title} {Berry curvature and the
  anomalous Hall effect in Heusler compounds},\ }\href@noop {} {\bibfield
  {journal} {\bibinfo  {journal} {Phys. Rev. B}\ }\textbf {\bibinfo
  {volume} {85}},\ \bibinfo {pages} {012405} (\bibinfo {year}
  {2012})}\BibitemShut {NoStop}%
\bibitem [{\citenamefont {Galanakis}\ \emph {et~al.}(2002)\citenamefont
  {Galanakis}, \citenamefont {Dederichs},\ and\ \citenamefont
  {Papanikolaou}}]{galanakis2002slater}%
  \BibitemOpen
  \bibfield  {author} {\bibinfo {author} {\bibfnamefont {I.}~\bibnamefont
  {Galanakis}}, \bibinfo {author} {\bibfnamefont {P.}~\bibnamefont
  {Dederichs}},\ and\ \bibinfo {author} {\bibfnamefont {N.}~\bibnamefont
  {Papanikolaou}},\ }\bibfield  {title} {\bibinfo {title} {Slater-Pauling
  behavior and origin of the half-metallicity of the full-Heusler alloys},\
  }\href@noop {} {\bibfield  {journal} {\bibinfo  {journal} {Phys. Rev.
  B}\ }\textbf {\bibinfo {volume} {66}},\ \bibinfo {pages} {174429} (\bibinfo
  {year} {2002})}\BibitemShut {NoStop}%
\bibitem [{\citenamefont {shukla}\ \emph {et~al.}(2020{\natexlab{a}})\citenamefont
  {shukla}, \citenamefont {kumar}}]{shukla2020destruction}%
  \BibitemOpen
  \bibfield  {author} {\bibinfo {author} {\bibfnamefont {V.}~\bibnamefont
  {Shukla}}, \bibinfo {author} {\bibfnamefont {S. O.}\ \bibnamefont {Kumar}},\ }\bibfield  {title} {\bibinfo
  {title} {Destruction of half-metallicity in Co$_2$VSn Heusler alloy due to X-Y swapping disorder},\ }\href@noop {} {\bibfield
  {journal} {\bibinfo  {journal} {J. Supercond. Nov. Magn.}\
  }\textbf {\bibinfo {volume} {33}},\ \bibinfo
  {pages} {3615} (\bibinfo {year}
  {2020}{\natexlab{a}})}\BibitemShut {NoStop}%
	\bibitem [{\citenamefont {Wang}\ \emph {et~al.}(2006)\citenamefont {Wang},
  \citenamefont {Yates}, \citenamefont {Souza},\ and\ \citenamefont
  {Vanderbilt}}]{PhysRevB.74.195118}%
  \BibitemOpen
  \bibfield  {author} {\bibinfo {author} {\bibfnamefont {X.}~\bibnamefont
  {Wang}}, \bibinfo {author} {\bibfnamefont {J.~R.}\ \bibnamefont {Yates}},
  \bibinfo {author} {\bibfnamefont {I.}~\bibnamefont {Souza}},\ and\ \bibinfo
  {author} {\bibfnamefont {D.}~\bibnamefont {Vanderbilt}},\ }\bibfield  {title}
  {\bibinfo {title} {Ab initio calculation of the anomalous Hall conductivity
  by Wannier interpolation},\ }\href
  {https://doi.org/10.1103/PhysRevB.74.195118} {\bibfield  {journal} {\bibinfo
  {journal} {Phys. Rev. B}\ }\textbf {\bibinfo {volume} {74}},\ \bibinfo
  {pages} {195118} (\bibinfo {year} {2006})}\BibitemShut {NoStop}%
\end{thebibliography}
\end{document}